\documentstyle[12pt]{article}
\newif\ifpic
\pictrue
\newcommand{\half}{{1\over2}}
\newcommand{\onarrow}[1]{\mathrel{\mathop{\longrightarrow}\limits^{#1}}}
\renewcommand{\H}{{\cal H}}
\newcommand{\Lm}{\Lambda}
\newcommand{\dn}{\hbox{\lower10pt\hbox{\Large$\searrow$}\kern-10pt $d_1$}}
\newcommand{\ov}{\overline}
\newcommand{\im}{\mathop {\rm Im}\nolimits}
\newcommand{\x}{{x^\mu}}
\renewcommand{\d}{{d}}
\renewcommand{\O}{\Omega}
\newcommand{\ds}{\displaystyle}
\newcommand{\p}{\partial}
\renewcommand{\a}{\alpha}
\newcommand{\dl}{\delta}
\newcommand{\lm}{\lambda}
\newcommand{\met}{\eta}
\newcommand{\ket}[1]{|#1\rangle}

\newcommand{\IR}{\rm I\!R}
\newcommand{\F}{{\cal F}}
\newcommand{\no}[1]{:\! #1 \!:}
\newcommand{\inbar}{\,\vrule height 1.5ex width .4pt depth0pt}
\newcommand{\IC}{\relax\hbox{$\inbar\kern-.3em{\rm C}$}}
\newcommand{\px}{\tilde x}
\newcommand{\hV}{\widehat{V}}
\newcommand{\hF}{\widehat\F}
\newcommand{\hQ}{\widehat{Q}}
\newcommand{\hE}{\widehat{E}}
\newcommand{\hL}{{\widehat{L}}}

\newcommand{\tV}{\widetilde{V}}

\newcommand{\tQ}{\widetilde{Q}}
\newcommand{\tE}{\widetilde{E}}

\newcommand{\Gr}{{\rm Gr}}
\newcommand{\tx}{\tilde x}
\newcommand{\onto}[1]{\stackrel{#1}{\longrightarrow}}
\newcommand{\fig}[1]{Fig.~\ref{fig:#1}}
\newcommand{\tab}[1]{Table~\ref{tab:#1}}
\newcommand{\eq}[1]{Eq.~(\ref{#1})}
\renewcommand{\arraystretch}{1.5}
\renewcommand{\theequation}{\thesection .\arabic{equation}}
\newcommand{\rH}{{\bf H}}
\newcommand{\rA}{{\bf A}}
\newcommand{\rS}{{\bf S}}
\newcommand{\rV}{{\bf V}}
\newtheorem{lemma}{Lemma}
\newtheorem{theorem}{Theorem}
\newcommand{\sev}[1]{%
  \setlength{\unitlength}{1cm}%
  \begin{picture}(1,0)(0,0)%
    \put(0,0){\vector(2,-1){1}}%
    \put(0.5,-0.1){\makebox(0,0)[b]{$#1$}}%
  \end{picture}%
}
\newcounter{nmb}
\newsavebox{\dts}        % for dotted line
\newsavebox{\dddt}       % first row of a shade
\newsavebox{\ddt}        % second row of a shade
\newsavebox{\shade}      % shade
\newsavebox{\sgrid}
\title{{}\hfill
    \raisebox{2cm}[0pt]{
      \begin{array}[b]{l}
        \small\mbox{MIT-CTP 2471}\\
        \small\tt hep-th/9511111
      \end{array}
    }\\
  String center of mass operator and its effect on BRST cohomology}
\author{Alexander Astashkevich\thanks{E-mail address: ast@mit.edu}\\
        \em Department of Mathematics\\
        MIT\\
        \and
        Alexander Belopolsky\thanks{Supported in part by funds provided
                         by the U.S.  Department of Energy (D.O.E.)  under
                         cooperative agreement \#DF-FC02-94ER40818.
                         E-mail address: belopols@mit.edu}\\
        \em Center for Theoretical Physics\\
        MIT}
\begin{document}
\edef\oldarraystretch{\arraystretch}
\renewcommand{\arraystretch}{0.8}
\maketitle
\renewcommand{\arraystretch}{\oldarraystretch}
\begin{abstract}
  We consider the theory of bosonic closed strings on the flat
  background $\IR^{25,1}$. We show how the BRST complex can be
  extended to a complex where the string center of mass operator,
  $x^\mu_0$, is well defined. We investigate the cohomology of the
  extended complex.  We demonstrate that this cohomology has a number
  of interesting features.  Unlike in the standard BRST cohomology,
  there is no doubling of physical states in the extended complex.
  The cohomology of the extended complex is more physical in a number
  of of aspects related to the zero-momentum states. In particular, we
  show that the ghost number one zero-momentum cohomology states are
  in one to one correspondence with the generators of the global
  symmetries of the background {\it i.e.}, the Poincar\'e algebra.
\end{abstract}
\newpage
\section{Introduction and summary}
As it was argued in ref.~\cite{bz95}, the gauge parameters that include
the zero mode of the $X^\mu$ operator have to be considered in order
to prove the complete dilaton theorems. If we allow $X^\mu$ to appear
in gauge parameters, it is natural to allow it to appear in the
physical states as well. This calls for an extended version of the
BRST complex where the zero mode of $X^\mu$ or, in other words, the
string center of mass operator $x_0^\mu$ is well defined.

In this paper we will consider the closed bosonic string with a flat
non-compact target space $\IR^{25,1}$. We will define the extended
complex simply as a tensor product of the BRST complex with the space
of polynomials of $D=26$ variables. Our main objective is to calculate
the semi-relative cohomology of this complex.

When we add new vectors to a complex, two phenomena may occur in
cohomology. First, some vectors that used to represent nontrivial
cohomology classes may become trivial, and second, some new cohomology
states may appear. Our original motivation \cite{bz95} to use the
extended complex was that the graviton trace ${\cal G}$, BRST-physical
state, was trivial in the extended complex. As we will see, the
extended complex provides many more examples of this kind. We will
show that only one out of $D^2+1$ ghost number two zero-momentum
BRST-physical states remains non-trivial in the extended complex. In
the BRST complex, the spectrum of non-zero momentum physical states is
doubled due to the presence of the ghost zero modes. We will see that
there is no such doubling in the extended complex: only ghost number
two states survive and all the ghost number three states become
trivial\footnote{The observation that ghost number three states are
  physical only in the case of finite space time volume was made in
  the book by Green, Schwarz, and Witten \cite{GSchW}. In the finite
  space time volume the momentum is discrete and $x_0^\mu=-i\p/\p
  p_\mu$ cannot be defined.}.  Returning to the second phenomenon, the
appearance of new physical states, we will be able to show that all
such states can be obtained from the old ones by differentiation with
respect to the continuous momentum parameter along the corresponding
mass shells. In this sense no new physical state will appear.

Zero momentum states will require special attention and the
calculation of the cohomology of the zero momentum extended complex is
technically the most difficult part of this work. We will find that
the ghost number one discrete states at zero momentum correctly
describe the global symmetries (Poincar\'e group) of the background.

The analysis presented in this paper shows that the semi-relative
cohomology of the extended complex correctly describes the physics of
the closed bosonic string around the flat $D=26$ background. The
arguments that lead to this conclusion are the following:
\begin{enumerate}
\item Ghost number two non-discrete physical states are the same as in
  the BRST complex up to infinitesimal Lorentz transformations. In
  this sense the extended complex is as good as BRST
  (section~\ref{s:lornz}).
\item There is no doubling of the physical states,---ghost number three
  BRST-physical states are trivial in the extended complex
  (section~\ref{s:nzclo}).
\item There is only one zero-momentum physical state at ghost number
  two which can be represented by the ghost dilaton
  (section~\ref{s:zclo} and ref.~\cite{bz95}).
\item Ghost number one discrete states are in one to one
  correspondence with the generators of the Poincar\'e group
  (section~\ref{s:zclo}).
\end{enumerate}

The paper is organized as follows. In section~\ref{s:extc} we start by
describing the extended complex and the nilpotent operator $\hQ$. We
define a cohomology problem for the extended complex and explain what
we are going to learn about its structure. In section~\ref{s:chi} we
formulate a simplified version of the problem in which we replace the
closed string BRST complex by its chiral part. Section~\ref{s:chi} is
devoted to a detailed analysis of the cohomology of this complex.  In
section~\ref{s:full} we investigate the (semi-relative) cohomology of
the full extended complex using the same methods as in
section~\ref{s:chi}. At the end of section~\ref{s:chi} and
section~\ref{s:full} we will formulate two theorems which summarize
our results on the structure of the cohomology of the chiral and the
full extended complexes. We present a detailed analysis of the Lorentz
group action on the cohomology of the extended complex in
section~\ref{s:lor}. For the sake of completeness we add two
appendices: \ref{app:spectral}, where we review some basic algebraic
facts which we use to calculate the cohomology and \ref{app:l}, where
we prove the results which are necessary to calculate the cohomology
at zero momentum.
\section{Extended complex}
\label{s:extc}
\setcounter{equation}{0} When we describe a string propagating in flat
uncompactified background, it seems natural to let the string center
of mass operator $x_0$ act on the state space of the theory. This
operator must satisfy the Heisenberg commutation relation
\begin{equation}
  \label{canon}
  [\a^\mu_n,x^\nu_0]=-i\,\dl_{n,0}\,\met^{\mu\nu},
\end{equation}
where $\met^{\mu\nu}$ is the Minkowski metric.  In order to
incorporate an operator with such properties we define an extended
Fock space as a tensor product of an ordinary Fock space with the
space of polynomials of $D$ variables:
\begin{equation}
  \label{extfock}
  \hF_p(\a,\ov\a)=\IC[x^0,\dots,x^{D-1}]\otimes\F_p(\a,\ov\a).
\end{equation}
All operators except $\a_0^\mu$ act only on the second
factor which is an ordinary Fock space, $x^\mu_0$ operators act by
multiplying the polynomials by the corresponding $x^\mu$ and, finally,
the action of $\a_0$ is defined by
\begin{equation}
  \label{a0act}
  \a^\mu_0= 1\otimes p^\mu
          - i\met^{\mu\nu}{\p\over\p x^\nu}\otimes1.
\end{equation}
Note that in the extended Fock module the action of $\a^\mu_0$ does
not reduce to the multiplication by $p^\mu$.

The extended Fock module is also a module over Virasoro algebra or,
strictly speaking, over a tensor product of two Virasoro algebras
corresponding to the left and right moving modes. The generators are
given by the usual formulae
\begin{equation}
  \label{virvir}
  \begin{array}{rcl}
      L_n&=&\ds{1\over2}\sum_m\met_{\mu\nu}\,\no{\a^\mu_m\,\a^\nu_{n-m}},\\
      {{\ov L}}_n&=&{1\over2}\sum_m
\met_{\mu\nu}\,\no{\ov\a^\mu_m\,\ov\a^\nu_{n-m}}.
  \end{array}
\end{equation}
The central charge of the extended Fock module is $26$, the same as
that of $\F_p(\a,\ov\a)$, and we can use it to construct a complex
with a nilpotent operator $\hQ$ (see refs.~\cite{GSchW,fms86}).
Following the standard procedure we define the extended complex as a
tensor product of the extended Fock module with the ghost module
$\F(b,c,\ov b,\ov c)$
\begin{equation}
  \label{extc}
  \hV_p=\F(b,c,\ov b,\ov c)\otimes\hF_p(\a,\ov\a)
\end{equation}
and introduce the nilpotent operator $\hQ$ as
\begin{equation}
  \label{hqdef}
    \hQ=\sum_n c_n \hL_{-n} -
  {1\over2}\sum_{m,n}(m-n)\no{c_{-m}c_{-n}b_{m+n}}+\mbox{a.h.},
\end{equation}
where we put a hat over Virasoro generators in order to emphasize that
they are acting on the extended Fock space.

We can alternatively describe the extended complex as a tensor product
of the BRST complex $V_p$ with the space of polynomials $\IC[x^0\cdots
x^{D-1}]$:
\begin{equation}
  \label{altdef}
  \hV_p=\IC[x^0\cdots x^{D-1}]\otimes V_p,
\end{equation}
and express the nilpotent operator $\hQ$ in terms of the BRST operator
$Q$ as follows
\begin{equation}
  \label{hqq}
\hQ=1\otimes Q-i{\p\over\p x^\mu}\otimes
                    \sum_n(c_n\a_{-n}^\mu+\ov c_n\ov\a_{-n}^\mu)-
            \Box\otimes c_0^+,
\end{equation}
where $c_0^+=(c_0+\ov c_0)/2$, and $\Box=\met^{\mu\nu}{\p^2\over \p
  x^\mu\p x^\mu}$.

So far we have constructed an extended complex $\hV_p=\IC[x^0\cdots
x^{D-1}]\otimes V_p$ equipped with a nilpotent operator $\hQ$ given by
\eq{hqq}. One can easily check that \eq{hqq} does define a nilpotent
operator using $Q^2=0$ and commutation relations between $\a^\mu_n$
operators.

\subsection{Cohomology of the extended complex}
\renewcommand{\theequation}{\thesubsection.\arabic{equation}}
\setcounter{equation}{0} Our major goal is to calculate the cohomology
of the operator $\hQ$ acting on the extended complex $\hV_p$ for
different values of the momentum $p$. More precisely, we will be
looking for the so-called semi-relative cohomology, which is the space
of vectors annihilated by $\hQ$ and $b_0^-=b_0-\ov b_0$ modulo the
image of $\hQ$ acting on the vectors annihilated by $b_0^-$ (see
refs.~\cite{nel89,dn91,zwiebachlong,bc94}).

In mathematical literature it is called the semi-infinite cohomology
of the ${\mit Vir\times Vir}$ algebra relative to the sub-algebra
${\cal L}_0^-$ generated by the central charge and $L_0^-=L_0-\ov L_0$
with the values in the extended Fock module $\hF_p$, and is denoted by
$\H({\mit Vir\times Vir},{\cal L}_0^-,\hF_p)$ (see
refs.~\cite{fey84,DBF}).  We denote this cohomology by
$H_S(\hQ,\hV_p)$.

Before we start the calculations, let us describe what kind of
information about the cohomology we want to obtain. Ordinarily, we are
looking for the dimensions of the cohomology spaces at each ghost
number. In the presence of the $x_0$ operator these spaces are likely
to be infinite dimensional (since multiplication by $x$ does not
change the ghost number). In order to extract reasonable information
about the cohomology we have to use an additional grading by the
degree of the polynomials in $x$.  The operator $\hQ$ mixes vectors of
different degrees and we can not a priori expect the cohomology states
to be represented by homogeneous polynomials.\footnote{For the $p=0$
  extended complex and only for this case we will be able to show that
  the cohomology can be represented by homogeneous polynomials, but
  this will come out as a non-trivial result.} Instead of the grading
on $\hV$ we have to use a decreasing filtration
\begin{equation}
  \label{hFFF}
  \cdots\supset F_{-k}\hV\supset F_{-k+1}\hV\supset\cdots\supset
             F_{0}\hV=V,
\end{equation}
where $F_{-k}\hV$ is the subspace of $\hV$ consisting of the vectors
with $x$ degree less or equal to $k$. The operator $\hQ$ respects this
filtration in a sense that it maps each subspace $F_r\hV$ to itself:
\begin{equation}
  \label{QF}
  \hQ:\; F_r\hV\to F_r\hV.
\end{equation}
This allows us to define a filtration on the cohomology of $\hQ$. By
definition $F_{-k}H(\hQ,\hV)$ consists of the cohomology classes which
can be represented by vectors with the $x$-degree less or equal to
$k$. Although as we mentioned above there is no $x$-grading on the
cohomology space, we can define a graded space which is closely
related to it. We define
\begin{equation}
  \label{graded}
\Gr_rH(\hQ,\hV) = F_{r}H(\hQ,\hV)/F_{r+1}H(\hQ,\hV).
\end{equation}
By definition $\Gr_rH(\hQ,\hV)$ (for $r\leq0$) consists of the
cohomology classes which can be represented by a vector with
$x$-degree $-r$ but not $-r+1$. These spaces carry a lot of
information about the cohomology structure. For example if we know the
dimensions of $\Gr_r$ for $r=0,-1,\dots,-k$ we can find the dimension
of $F_{-k}$ as their sum. On the other hand the knowledge of
representatives of $\Gr_r$ states is not enough to find the
representatives of cohomology classes. This is so because $\Gr_r$
spaces contain information only about the leading in $x$ terms of
the cocycles of $\hQ$.

To find the graded space $\Gr H=\bigoplus\Gr_rH$ we will use the
machinery known as the method of spectral sequences. The idea is to
build a sequence of complexes $E_n$ with the differentials $d_n$ such
that $E_{n+1}=H(d_n,E_n)$ which converges to the graded space $\Gr H$.
In our case we will be able to show that all differentials $d_n$ for
$n>2$ vanish and thus $E_3=E_4=\cdots=E_\infty$. Therefore, we will
never have to calculate higher then the third terms in the spectral
sequence. We will give some more details on the application of the
method of spectral sequences to our case in \ref{app:spectral}.
\section{Chiral extended complex}
\label{s:chi}
\renewcommand{\theequation}{\thesection.\arabic{equation}}
\setcounter{equation}{0} Before attempting a calculation of the
cohomology of the full extended complex let us consider its chiral
version. This is a warm up problem which, nevertheless, captures the
major features.

We replace the Fock space $\F_p(\a,\ov\a)$ by its chiral version,
$\F_p(\a)$ which is generated from the vacuum by the left moving modes
$\a_n$ only.

Repeating the arguments of the previous section we conclude that the
chiral version of $\hQ$ is given by
\begin{equation}
  \label{xbrst}
  \hQ =
  1\otimes Q-i{\p\over\p x^\mu}\otimes \sum_nc_n\a^\mu_{-n}-
  {1\over2}\met^{\mu\nu}\Box\otimes c_0.
\end{equation}
We will calculate the cohomology of chiral extended complex $\hV_p$
for three different cases: case $p^2\neq0$, which describes the
massive spectrum; case $p^2=0$---the massless one; and case
$p\equiv0$, which besides the particular states from the massless
spectrum
describes a number of discrete states.
\subsection{Massive states }
\renewcommand{\theequation}{\thesubsection.\arabic{equation}}
\setcounter{equation}{0} Let us start the calculation of the
cohomology of the chiral extended complex $\hV_p$ by considering the
case of $p^2\neq0$. For this case the cohomology of the BRST complex
is non-zero only for ghost number one and ghost number two.  The
cohomology contains the same number of ghost number one and two states
which can be written in terms of dimension one primary matter states.
Let $\ket{v,p}\in V_p$ be a dimension one primary state with no ghost
excitations; then the following states,
\begin{equation}
  \label{reps}
    c_0c_1\ket{v,p}\qquad\mbox{and}\qquad
    c_1\ket{v,p},
\end{equation}
represent nontrivial cohomology classes and, moreover, each cohomology
class has a representative of this kind (see ref.~\cite{fgz86}).

We will calculate the cohomology of the extended complex in two steps.
First, we extend the BRST complex by adding polynomials of one variable
$\px=(p\cdot x)$. The resulting space,
\begin{equation}
  \label{px}
  \tV_p = \IC[\px]\otimes V_p,
\end{equation}
is a subcomplex of $\hV_p$ and we define its cohomology as $H(\tQ,
\tV_p)$, where $\tQ$ is the restriction of $\hQ$ on
$\tV_p$. Calculation of $\Gr H(\tQ,\tV_p)$ is the objective of the
first step. Second, we obtain the full extended space as
a tensor product of $\tV_p$ with the polynomials of the transverse
variables
\begin{equation}
  \label{pxx}
  \hV_p = \IC[\tx^1\dots,\tx^{D-1}]\otimes \tV_p,
\end{equation}
where
\begin{equation}
  \label{tradef}
   \tx^i=x^i-{p^i(p\cdot x)\over p^2}.
\end{equation}
Using $\Gr H(\tQ,\tV_p)$ found in the first step, we will calculate
$\Gr H(\hQ,\hV_p)$.

Let us calculate  $\Gr H(\tQ,\tV_p)$.
\begin{figure}[t]
  \begin{center}
    \leavevmode
    \setlength{\unitlength}{35pt}
    \begin{picture}(7,6)(-0.5,-0.5)
      \ifpic
      \savebox{\dts}(1.4,0){%
        \begin{picture}(1.4,0)
          \multiput(.2,0)(0.14,0){10}{\makebox(0,0){$\cdot$}}
        \end{picture}
        }%
                                %vertical lines
      \multiput(0,-0.5)(1,0){6}{\line(0,1){6}}
      \setcounter{nmb}{-5}
      \multiput(0,-1)(1,0){5}{\addtocounter{nmb}{1}%
        \makebox(1,0.5)[c]{\small$r\!=\!\!\arabic{nmb}$}}
      \put(5.05,-0.5){\line(0,1){6}}%
                                %horisontal lines
      \multiput(-0.5,0)(0,1){6}{\line(1,0){5.5}}
      \setcounter{nmb}{-1}
      \multiput(-0.3, 0)(0,1){5}{\addtocounter{nmb}{1}%
        \makebox(0,1)[cr]{\small$s=\arabic{nmb}$}}%
                                %diagonal lines
      \multiput(-0.5,5.5)(0.1,-0.1){51}{\makebox(0,0){$\cdot$}}
      \multiput( 0.5,5.5)(0.1,-0.1){41}{\makebox(0,0){$\cdot$}}
      \multiput( 1.5,5.5)(0.1,-0.1){31}{\makebox(0,0){$\cdot$}}
      \multiput( 2.5,5.5)(0.1,-0.1){21}{\makebox(0,0){$\cdot$}}
      \multiput( 3.5,5.5)(0.1,-0.1){11}{\makebox(0,0){$\cdot$}}
      \setcounter{nmb}{-1}
      \multiput(4.5,.5)(0,1){5}{\addtocounter{nmb}{1}%
        \usebox{\dts}\makebox(0,0)[l]{$\:{\rm Gh\#}=\arabic{nmb}$}
        }
      \put(2.5,2.5){\vector(0,1){1}\makebox(0,0.8)[tr]{\small$\partial_0$}}
      \put(2.5,2.5){\vector(1,0){1}}
      \put(2.1,2.5){\makebox(1,0.3)[rt]{\small$\partial_1$}}
      \put(2.5,2.5){\vector(2,-1){2}}
      \put(2.8,1.6){\makebox(1,0)[rb]{\small$\partial_2$}}
      \fi
    \end{picture}
  \end{center}
  \caption{Anatomy of a double complex}
  \label{fig:dbl}
\end{figure}
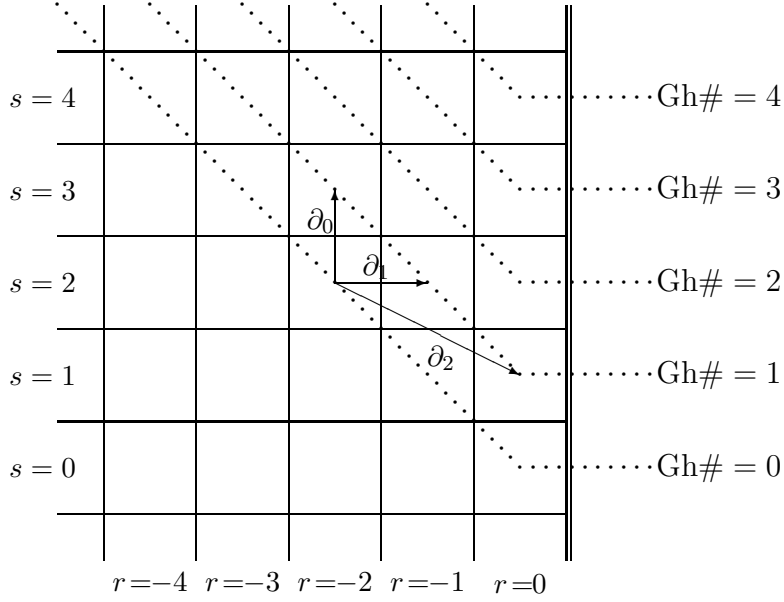
Beside the ghost number, complex $\tV_p$ has an additional
grading---the $x$-degree. According to these two gradings we can write
$\tV_p$ as a double sum
\begin{equation}
  \label{dsum}
   \tV_p=\bigoplus_{r,s}\tE_0^{r,s}(p),
\end{equation}
where $\tE_0^{r,s}=\px^{-r}\otimes V_p^{(r+s)}$ is the space of ghost
number $r+s$ states with $-r$ factors of $\px$. Note that in our
notations $r\leq0$.

It will be convenient to represent a double graded complex like
$\tV_p$ graphically by a lattice (see~Fig.~\ref{fig:dbl}) where each
cell represents a space $\tE_0^{r,s}$, columns represent the spaces
with definite $x$-degrees and the diagonals represent the spaces with
definite ghost numbers.

The action of $\tQ$ on $\tV_p$ can be easily derived from the general
formu\-la~(\ref{xbrst}). Any vector from $\tE_0^{-k, s}$ can be
represented as $\px^k\otimes\ket{v,p}$, where $\ket{v,p}\in
V^{(s-k)}_p$ is a vector from the BRST complex $V_p$ with ghost number
$s-k$. Applying $\tQ$ to this state we obtain
\begin{equation}
  \label{pxq}
  \begin{array}{rcl}\ds
   \tQ\;  \px^k\otimes \ket{v,p} &= &\px^k\otimes Q\ket{v,p} \\[6pt]
    &-&\ds ik\,\px^{k-1}\otimes \sum_n c_n\,(p\cdot\a_{-n})\,\ket{v,p}\\
    &-&\ds {k\,(k-1)\over2}\,p^2\,\px^{k-2}\otimes c_0\,\ket{v,p}\\
  \end{array}
\end{equation}
According to \eq{pxq} we decompose $\tQ$ in the sum of operators with
a definite $x$-degree
\begin{equation}
  \label{ddd}
  \tQ=\p_0+\p_1+\p_2,
\end{equation}
where each $\p_n$ reduces the $x$-degree, or increases $r,$ by $n$ (see
Fig.~\ref{fig:dbl}).

Now we start building the spectral sequence of the complex $(\tV_p,
\tQ)$. For a short review of the method see \ref{app:spectral}. The
first step, the calculation of $\tE_1^{r,s}=\Gr_rH^{r+s}(\p_0,\tV)$,
reduces to the calculation of the cohomology of the BRST complex.
Indeed, according to \eq{pxq}, $\p_0=1\otimes Q$, and therefore
\begin{equation}
  \label{e1}
  \tE_1^{r,s}=\px^{-r}\otimes H^{(r+s)}(Q,V_p).
\end{equation}
\setlength{\unitlength}{30pt}
\newcommand{\dt}{\makebox(0,0){{\tiny$\cdot$}}}
\savebox{\dddt}{%
  \begin{picture}(1,0)%
    \multiput(0,0)(0.1,0){11}{\dt}
  \end{picture}
  }
\savebox{\ddt}{%
  \begin{picture}(1,0)%
    \multiput(0.05,0)(0.1,0){10}{\dt}
  \end{picture}
  }
\savebox{\shade}{%
  \begin{picture}(1,1)
    \multiput(0,0)(0,0.1){11}{\usebox{\dddt}}
    \multiput(0,0.05)(0,0.1){10}{\usebox{\ddt}}
  \end{picture}
}
\savebox{\sgrid}{%
  \begin{picture}(5,5)
    \ifpic
    \multiput(0,-0.5)(1,0){4}{\line(0,1){4}}
    \multiput(-0.5,0)(0,1){4}{\line(1,0){3.5}}
    \put(3.05,-0.5){\line(0,1){4}}
    \setcounter{nmb}{0}
    \multiput(-0.7, 0)(0,1){3}{\addtocounter{nmb}{1}%
      \makebox(0,1){\small$\arabic{nmb}$}}
    \put(-0.7,3){\vector(0,1){0.5}}
    \put(-0.7,3.6){\makebox(0,0)[b]{$s$}}
    \setcounter{nmb}{-3}
    \multiput(0, -0.7)(1,0){3}{\addtocounter{nmb}{1}%
      \makebox(1,0){\small$\arabic{nmb}$}}
    \put(3,-0.7){\vector(1,0){0.5}}
    \put(3.6,-0.7){\makebox(0,0)[l]{$r$}}
    \fi
  \end{picture}
  }
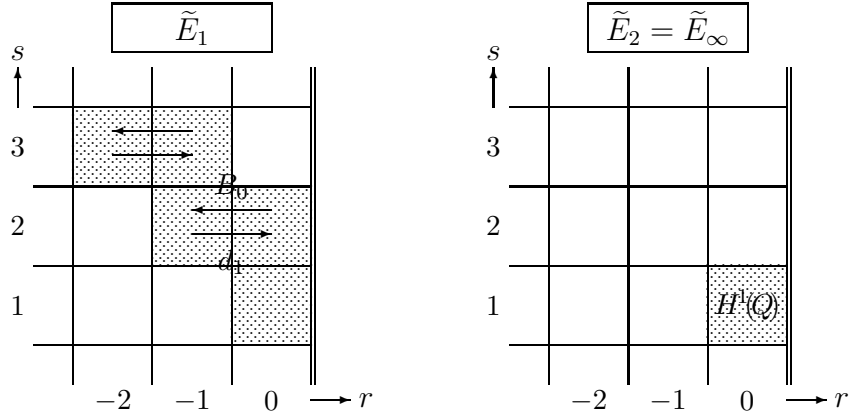
\begin{figure}[t]
  \begin{center}
    \leavevmode
    \begin{picture}(5,5.5)(-1,-1)
      \ifpic
      \put(0,0){\usebox{\sgrid}}
      \multiput(2,1)(-1,1){2}{\usebox{\shade}}
      \multiput(2,0)(-1,1){3}{\usebox{\shade}}
      \multiput(1.5,1.4)(-1,1){2}{\vector(1,0){1}}
      \multiput(2.5,1.7)(-1,1){2}{\vector(-1,0){1}}
      \put(2,1.9){\makebox(0,0)[b]{\small$B_0$}}
      \put(2,1.2){\makebox(0,0)[t]{\small$d_1$}}
      \put(0.5,3.7){\framebox(2,.6){{$\tE_1$}}}
      \fi
    \end{picture}\hspace\unitlength
    \begin{picture}(5,5.5)(-1,-1)
      \ifpic
      \put(0,0){\usebox{\sgrid}}
      \put(2,0){\usebox{\shade}}
      \put(2,0){\makebox(1,1){\small$H^{\!1}\!(\!Q\!)$}}
      \put(0.5,3.7){\framebox(2,.6){$\tE_2=\tE_\infty$}}
      \fi
    \end{picture}
  \end{center}
  \caption{Spectral sequence for $\widetilde V_p$}
  \label{fig:tvp}
\end{figure}
As we mentioned above, the BRST complex has nontrivial cohomology
only at ghost numbers one and two. Thus the space $\tE_1=\bigoplus
\tE^{r,s}_1$ looks as shown in Fig.~\ref{fig:tvp} (left), where shaded
cells correspond to non-zero spaces.

The differential $d_1$ is induced on $\tE_1$ by $\p_1$, and acts from
$\tE_1^{r,s}$ to $\tE_1^{r+1,s}$ as shown in Fig.~\ref{fig:tvp}
(left). Since there are no states below the ghost number one and above
ghost number two, the cohomology of $d_1$ at ghost number one is
given by its kernel:
\begin{equation}
  \label{kerd1}
  \tE_2^{r,1-r}=\ker d_1,
\end{equation}
and at ghost number two by the quotient of $\tE_2^{r,2-r}$ by the
image of $d_1$:
\begin{equation}
  \label{fac}
  \tE_1^{r,2-r}= \tE_2^{r,2-r}/\im d_1.
\end{equation}
We are going to show that $d_1$ establishes an isomorphism of the
corresponding spaces and, therefore, the only non-empty component of
$\tE_2$ is $\tE_2^{0,1}\simeq H^1(Q,V_p)$ as shown in
\fig{tvp} (right).

Consider an operator $B_0=\px\otimes b_0$. This operator is well
defined on $\tE_1$ {\it i.e.}, it maps cohomology classes to cohomology
classes. On the other hand, its anticommutator with $d_1$ is given by
\begin{equation}
  \label{p1b0}
  \{d_1, B_0\} =p^2 \,\hat k,
\end{equation}
where $\hat k$ the $x$-degree operator. The last equation shows that
if $p^2\neq0$, nontrivial cohomology of $d_1$ may exist only in $\hat
k =0$ subspace of $\tE_2$. Moreover, if we apply $\{d_1,
B_0\}=d_1\,B_0+B_0\,d_1$ to ghost number one states only the second
term will survive because there are no ghost number zero states in
$\tE_1$. Thus we conclude that up to a diagonal matrix $B_0$ is an
inverse operator to $d_1$. Since $\tE_1^{r,1-r}$ and $\tE_1^{r+1,1-r}$
have the same dimension and $d_1$ is invertible it is an isomorphism
between $\tE_2^{r,1-r}$ and $\tE_2^{r+1,1-r}$ for any $r\leq0$.

As shown in \fig{tvp} (right), $\tE_2$ contains only one non-empty
component. This means that second differential $d_2$ and all higher are
necessarily zero and the spectral sequence collapses at
$\tE_2=\tE_\infty$. Therefore, we conclude that
\begin{equation}
  \label{cohtq}
  \Gr H(\tQ,\tV_p)=H^1(Q,V_p).
\end{equation}

The second step in our program is trivial because the spectral
sequence $\{\hE_n\}$ of the full complex
\begin{equation}
  \label{hvtv}
  \hV_p=\IC[\tx^1,\dots,\tx^{D-1}]\otimes\tV_p,
\end{equation}
stabilizes at $\hE_1$ and
\begin{equation}
  \label{hvcoh}
 \hE_1=\hE_\infty=\IC[\tx_1,\dots,\tx_{D-1}]\otimes\Gr H(\tQ,\tV_p).
\end{equation}
This happens simply because, according to \eq{cohtq}, $\Gr
H(\tQ,\tV_p)$, and thus $\hE_1$, contains only ghost number one states
and therefore $d_1$ and all higher differentials must
vanish. Combining Eqs.~(\ref{hvcoh}) and~(\ref{cohtq}) we obtain
\begin{equation}
  \label{cohhq}
   \Gr H(\hQ,\hV_p)=\IC[\tx^1,\dots,\tx^{D-1}]\otimes H^1(Q,V_p).
\end{equation}
This completes our analysis of the cohomology of the chiral extended
complex for $p^2\neq0$.

\subsection{Massless states}
\renewcommand{\theequation}{\thesubsection.\arabic{equation}}
\setcounter{equation}{0} The analysis presented above can not be
applied to the light-cone, $p^2=0$.  We could, in principle, repeat all
the arguments using $\xi\cdot x$ instead of $\px$, where $\xi$ is some
vector for which $\xi\cdot p\neq0$, to build $\tV$ and this would work
everywhere except at the origin of the momentum space, $p=0$. Yet it is
instructive to make a covariant calculation in this case.  Since there
is no covariant way to choose a vector $\xi$ we can not apply our two
step program. Instead we will start from scratch and build a spectral
sequence for the whole module
\begin{equation}
  \label{hv0}
  \hV_p=\IC[x^0, \dots,x^{{D-1}}]\otimes V_p,
\end{equation}
graded by the total $x$-degree.

According to Eq.~(\ref{xbrst}), we can decompose $\hQ$ into a sum of
operators of definite $x$-degree
\begin{equation}
  \label{ddn}
  \hQ=\p_0+\p_1+\p_2,
\end{equation}
where
\begin{equation}
  \label{dxpr}
  \begin{array}{rcl}
    \p_0&=&1\otimes Q,\\
    \p_1&=&\ds-i{\p\over\p x^\mu}\otimes \sum_nc_n\a_{-n}^\mu,\\
    \p_2&=&\ds-\half\Box\otimes c_0.\\
  \end{array}
\end{equation}

The first step is to find cohomology of $\p_0$, which is just the tensor
product of the BRST cohomology $H(Q,V_p)$ with the space of polynomials
\begin{equation}
  \label{e1p0}
  E_1=H(\p_0,\hV_p)=\IC[x^0\cdots x^{D-1}]\otimes H(Q,V_p).
\end{equation}
Multiplying the representatives of $ H(Q,V_p)$ by arbitrary
polynomials in $x$ we obtain the following representatives of $E_1$
cohomology classes
\begin{equation}
  \label{mlrep}
  \begin{array}{l}
    P_\mu(x)\otimes c_1\a_{-1}^\mu\ket{p},\\
    Q_\mu(x)\otimes c_0 c_1\a_{-1}^\mu\ket{p},\\
  \end{array}
\end{equation}
where $P_\mu(x)$ and $Q_\mu(x)$ are polynomials in $x$ that satisfy
the transversality condition, $p^\mu Q_\mu(x)=p^\mu P_\mu(x)=0$, and
are not proportional to $p_\mu$. These transversality conditions come
from the same conditions on BRST cohomology classes at $p^2=0$. The
first differential acts non-trivially from ghost number one to ghost
number two states according to the following formula
\begin{equation}
  \label{d1ac}
  d_1:\quad P_\mu\to Q_\mu=-ip^\nu{\p\over\p x^\nu}P_\mu.
\end{equation}
It is easy to check that the map (\ref{d1ac}) is surjective and
therefore $E_2^{r,s}=0$ for $r+s=2$. As expected the cohomology of the
massless complex has a similar structure to that of the massive one.
There are no cohomology states with ghost number two and there is an
infinite tower of ghost number one states with different $x$-degree.

\subsection{Cohomology of the zero momentum chiral complex}
\renewcommand{\theequation}{\thesubsection.\arabic{equation}}
\setcounter{equation}{0} The zero momentum complex is exceptional.
Already in the BRST cohomology we encounter additional ``discrete''
states at exotic ghost numbers (see ref.~\cite{dn92}). The cohomology
is one dimensional at ghost numbers zero and three and $D$-dimensional
at ghost numbers one and two. Explicit representatives for these
classes can be written as given in \tab{zreps}.
\begin{table}[t]
  \begin{center}
    \leavevmode
  \begin{tabular}[c]{|c|c|c|}
    \hline
    Ghost \#&Representatives&Dimension\\
    \hline
    \hline
    $3$&$c_1c_0c_{-1}\ket0$&$1$\\
    \hline
    $2$&$c_0\a^\mu_{-1}\ket0$&$D$\\
    \hline
    $1$&$   \a^\mu_{-1}\ket0$&$D$\\
    \hline
    $0$&$\ket0$&$1$\\
    \hline
  \end{tabular}
  \end{center}
\caption{Chiral BRST cohomology at $p=0$}
  \label{tab:zreps}
\end{table}

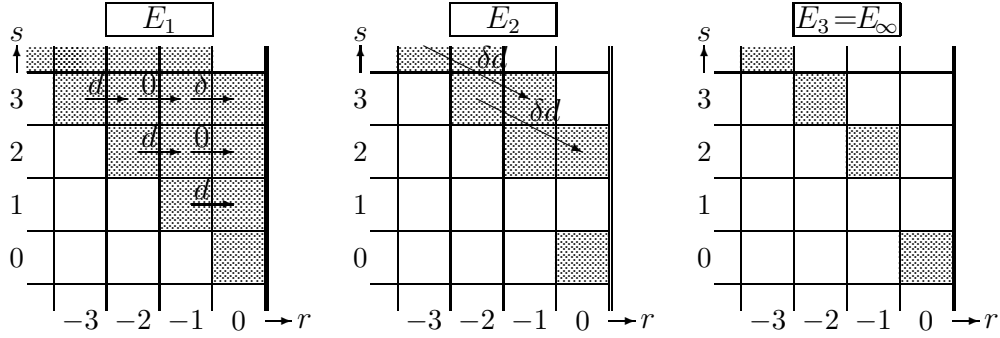
\begin{figure}[t]
\setlength{\unitlength}{20pt}
\savebox{\dddt}{%
  \begin{picture}(1,0)%
    \multiput(0,0)(0.1,0){11}{\dt}
  \end{picture}
  }
\savebox{\ddt}{%
  \begin{picture}(1,0)%
    \multiput(0.05,0)(0.1,0){10}{\dt}
  \end{picture}
  }
\savebox{\shade}{%
  \begin{picture}(1,1)
    \multiput(0,0)(0,0.1){11}{\usebox{\dddt}}
    \multiput(0,0.05)(0,0.1){10}{\usebox{\ddt}}
  \end{picture}
}
\newsavebox{\halfshade}
\savebox{\halfshade}{%
  \begin{picture}(1,1)
    \multiput(0,0)(0,0.1){5}{\usebox{\dddt}}
    \multiput(0,0.05)(0,0.1){5}{\usebox{\ddt}}
  \end{picture}
}
\savebox{\sgrid}{%
  \begin{picture}(5,5)
    \ifpic
    \multiput(0,-0.5)(1,0){5}{\line(0,1){5}}
    \multiput(-0.5,0)(0,1){5}{\line(1,0){4.5}}
    \put(4.05,-0.5){\line(0,1){5}}
    \setcounter{nmb}{-1}
    \multiput(-0.7, 0)(0,1){4}{\addtocounter{nmb}{1}%
      \makebox(0,1){\small$\arabic{nmb}$}}
    \put(-0.7,4){\vector(0,1){0.5}}
    \put(-0.7,4.6){\makebox(0,0)[b]{$s$}}
    \setcounter{nmb}{-4}
    \multiput(0, -0.7)(1,0){4}{\addtocounter{nmb}{1}%
      \makebox(1,0){\small$\arabic{nmb}$}}
    \put(4,-0.7){\vector(1,0){0.5}}
    \put(4.6,-0.7){\makebox(0,0)[l]{$r$}}
    \fi
  \end{picture}
  }
  \begin{center}
    \leavevmode
    \begin{picture}(5.5,6.5)(-1,-1)
      \ifpic
      \put(0,0){\usebox{\sgrid}}
      \put(-0.5,4){\usebox{\halfshade}}
      \multiput(0,4)(1,0){3}{\usebox{\halfshade}}
      \put(3,3){\usebox{\shade}}
      \multiput(3,2)(-1,1){2}{\usebox{\shade}}
      \multiput(3,1)(-1,1){3}{\usebox{\shade}}
      \multiput(3,0)(-1,1){4}{\usebox{\shade}}
      \multiput(0.8,3.6)(1,-1){3}{\makebox(0,0)[b]{$d$}}
      \multiput(0.6,3.5)(1,-1){3}{\vector(1,0){0.8}}
      \multiput(1.8,3.6)(1,-1){2}{\makebox(0,0)[b]{$0$}}
      \multiput(1.6,3.5)(1,-1){2}{\vector(1,0){0.8}}
      \put(2.8,3.6){\makebox(0,0)[b]{$\delta$}}
      \put(2.6,3.5){\vector(1,0){0.8}}
      \put(1,4.7){\framebox(2,.6){{$E_1$}}}
      \fi
    \end{picture}\hspace\unitlength
    \begin{picture}(5.5,6.5)(-1,-1)
      \ifpic
      \put(0,0){\usebox{\sgrid}}
      \multiput(0,4)(1,0){2}{\usebox{\halfshade}}
      \multiput(3,2)(-1,1){2}{\usebox{\shade}}
      \multiput(2,2)(-1,1){2}{\usebox{\shade}}
      \put(3,0){\usebox{\shade}}
      \multiput(1.5,3.5)(-1,1){2}{\vector(2,-1){2}}
      \multiput(2.5,3.3)(-1,1){2}{\makebox(0,0)[l]{$\delta d$}}
      \put(1,4.7){\framebox(2,.6){{$E_2$}}}
      \fi
    \end{picture}\hspace\unitlength
    \begin{picture}(5.5,6.5)(-1,-1)
      \ifpic
      \put(0,0){\usebox{\sgrid}}
      \put(0,4){\usebox{\halfshade}}
      \multiput(2,2)(-1,1){2}{\usebox{\shade}}
      \put(3,0){\usebox{\shade}}
      \put(1,4.7){\framebox(2,.6){{$E_3\!=\!\!E_{\!\infty}$}}}
      \fi
    \end{picture}
  \end{center}
  \caption{Spectral sequence for $\widehat V_0$}
  \label{fig:spt2}
\end{figure}

Let us denote the direct sum of spaces $E_n^{r,s}$ with the same ghost
number $m=r+s$ by $E_n^{(m)}$:
\begin{equation}
  \label{adef}
  E_n^{(m)}\equiv\bigoplus_{r\leq0}E_n^{r,m-r}
\end{equation}
As usual, the cohomology of $\p_0$ can be written in terms of the BRST
cohomology as
$$
E_1^{(m)}=H^m(\p_0,\hV_0)=\IC[x^0,\dots,x^{{D-1}}]\otimes H^m(Q,V_0).
$$
According to \tab{zreps}, $H^0(Q,V_0)\simeq H^3(Q,V_0)\simeq\IC$.
Therefore, $E_1^{(0)}$ and $E_1^{(3)}$ are isomorphic to the space of
polynomials $\IC[\x]$. Similarly, $E_1^{(1)}$ and $E_1^{(2)}$ are
isomorphic to the space of polynomial vector fields $\IC^{D}[\x]$.
Evaluating the action of $\p_1$ on the representatives (see
\tab{zreps}), we get the following sequence
\begin{equation}
  \label{seq}
   0\to\IC[\x]\stackrel{\nabla}{\to}\IC^{D}[\x]
      \stackrel{0}{\to}\IC^{D}[\x]\stackrel{\nabla}{\to}\IC[\x]\to0,
\end{equation}
where the first nabla-operator is the gradient, which maps scalars to
vectors and the second is the divergence, which does the opposite (and we
have dropped an insignificant factor $-i$).  It is convenient to
interpret $\IC[\x]=\O^0$ and $\IC^{D}[\x]=\O^1$ as a space of
polynomial zero and one-forms on the Minkowski space.  The first
nabla-operator in \eq{seq} will be interpreted as an exterior
derivative $\d$ and the second as its Hodge conjugate $\dl$. With new
notation we rewrite the sequence in \eq{seq} as
\begin{equation}
  \label{hseq}
  0\to\O^0\stackrel{\d}{\to}\O^1\stackrel{0}{\to}
                            \O^1\stackrel{\dl}{\to}\O^0\to0.
\end{equation}
This is illustrated by \fig{spt2} (left).
Note that individual cells in \fig{spt2} correspond to the subspaces
of homogeneous polynomials rather than whole $\O^1$ or $\O^0$. One can
easily calculate cohomology of this complex and obtain
\begin{equation}
  \label{e2}
  \begin{array}{rcl}
  E_2^{(0)}&=&\IC,\\
  E_2^{(1)}&=&\O^1/\d\O^0,\\
  E_2^{(2)}&=&\ker(\dl,\O^1),\\
  E_2^{(3)}&=&0.
  \end{array}
\end{equation}
Thus the second term in the spectral sequence has the structure
presented by \fig{spt2} (middle).

Now we have to calculate $d_2$. This differential acts as shown in
\fig{spt2} (middle) and can be found from the following formula (see
\ref{app:spectral}):
\begin{equation}
  \label{d2}
  d_2=\p_2-\p_1\p_0^{-1}\p_1.
\end{equation}
Let $P=P_\mu\d x^\mu\in\O^1$ be a polynomial one form. Corresponding
state representing a $E_2^{(1)}$ cohomology class is given by
\begin{equation}
  \label{arep}
  P_\mu\otimes  c_1\a_{-1}^\mu\ket{0}.
\end{equation}
Applying $\p_1$ to (\ref{arep}) we obtain
\begin{equation}
  \label{d1act}
  \p_1\;P_\mu\otimes \a_{-1}^\mu c_1\ket{0} =
  -i{ \p P_\mu\over\p x_\mu}\otimes   c_{-1} c_1\ket{0}=
    \p_0\;i\half{\p P_\mu\over\p x_\mu}\otimes  c_0\ket{0},
\end{equation}
where we use the metric $\met^{\mu\nu}$ to raise and lower indices. From
\eq{d1act} we derive that
\begin{equation}
  \label{dd0d}
  \p_1\p_0^{-1}\p_1 \; P_\mu\otimes \a_{-1}^\mu c_1\ket{0}=
  \p_1\;{i\over2}{\p P_\mu\over\p x_\mu}\otimes c_0\ket{0}=
  \half{\p^2 P_\mu\over\p x_\mu\p x^\nu }\otimes\a_{-1}^\nu c_1c_0\ket{0}.
\end{equation}
With the definition of $\p_2$ (see \eq{dxpr}) we immediately get
\begin{equation}
  \label{p2act}
  \p_2 \; P_\mu\otimes \a_{-1}^\mu c_1\ket{0}=
    \half\Box P_\mu c_1c_0\ket{0}.
\end{equation}
By adding the last two equations, and dropping an insignificant factor
$1/2$, we see that the second differential acts on
$E_2^{(1)}=\IC^{D}[\x]/\nabla\IC[\x]$ as the following matrix
differential operator:
\begin{equation}
  \label{d2fin}
  (d_2)_\mu^\nu=\Box\dl_\mu^\nu-\p_\mu\p^\nu,
\end{equation}
or, using the differential form interpretation $E_2^{(1)}=\O^1/\d\O^0$,
\begin{equation}
  \label{d2om}
  d_2=\Box-\d\dl=\dl\d.
\end{equation}
In order to calculate $E_3=H(d_2,E_2)$ we have to calculate the
cohomology of the following complex:
\begin{equation}
  \label{e3df}
  0\to\IC\stackrel{0}{\to}\O^1/\d\O^0\stackrel{\dl\d}{\to}\ker(\dl,\O^1)\to0,
\end{equation}
or equivalently
\begin{equation}
  \label{2in1}
   0\to\O^0\stackrel{\d}{\to}\O^1\stackrel{\dl\d}{\to}
                             \O^1\stackrel{\dl}{\to}\O^0\to0.
\end{equation}
The later sequence is known to be exact in the last two
terms\footnote{We would like to thank Jeffrey Goldstone for pointing
  this out to us.} and thus $E_3$ is given by
\begin{equation}
  \label{e3}
  \begin{array}{rcl}
  E_3^{(0)}&=&\IC,\\
  E_3^{(1)}&=&\ker(\dl\d,\O^1)/\d\O^0,\\
  E_3^{(2)}&=&E_3^{(3)}\;=\;0.
  \end{array}
\end{equation}
Looking at \fig{spt2} (right) one can easily deduce that all
differentials $d_k$ for $k\geq3$ vanish. This allows us to conclude
that
\begin{equation}
  \label{p0answer}
\Gr_rH^{r+s}(\hQ,\hV_0)=E^{r,s}_\infty=E_3^{r,s},
\end{equation}
and $\Gr H$ has the structure shown in \fig{spt2} (right). Since
cohomology is non-trivial only at one ghost number for every
$x$-degree, we can easily find the dimensions of different $\Gr_rH$
spaces. We present the summary of our results on the cohomology of the
chiral extended complex in the following theorem.
\begin{theorem}
  The cohomology of the chiral extended complex $\hV_p$ admits a
  natural filtration (by the {\em minus\/} $x$-degree) $\F_r H(\hV_p,\hQ)$.
  The cohomology can be described using the associated graded spaces
  $$\Gr_r H(\hV_p,\hQ)=\F_r H(\hV_p,\hQ)/\F_{r+1} H(\hV_p,\hQ)$$ as follows
  \begin{enumerate}
  \item Non-zero momentum ($p\not\equiv0$): cohomology is trivial unless
    $p^2=n-1$, where $n$ is a non-negative integer. In the latter
    case
    \begin{eqnarray*}
      \dim \Gr_r H^1(\hV_p,\hQ)&=&\ds{24-r\choose 24}\,d_n,\\
      \dim H^l(\hV_p,\hQ)&=&0\qquad\mbox{for $l\neq1$},
    \end{eqnarray*}
    where $d_n$ denotes the number of the BRST states at mass level
    $n$. These numbers are generated by
    $$\prod_{k=1}^\infty(1-z^k)^{-24}=\sum_{n=1}^\infty d_n z^n.$$
    see ref.~\cite{GSchW}.
  \item Zero momentum ($p\equiv0$): the cohomology appears at ghost
    number zero (the vacuum)
    \begin{displaymath}
            \dim H^0(\hV_p,\hQ)=1,
    \end{displaymath}
    and ghost number one (physical states, recall that $r$ is {\em
      minus\/} $x$-degree).
    \begin{displaymath}
      \dim \Gr_r H^1(\hV_p,\hQ)=\ds\cases{
        \hfill0\hfill& if $r=0$;\cr
        \ds\hfill{D(D-1)\over2}\hfill&for $r=-1$;\cr
        \ds{D(D-2)(D+2)\over3}\hfill&for $r=-2$;\cr
        \hfill\chi_r\hfill&for $r\leq-3$.}
    \end{displaymath}
    where
    $$\chi_r=D\,{D-1-r\choose D-1}+{D-4-r\choose D-1}-
    {D-r\choose D-1}- D\,{D-3-r\choose D-1}.$$
    The cohomology is trivial for all the other ghost numbers
    \begin{displaymath}
      \dim H^l(\hV_p,\hQ)=0\qquad\mbox{for $l\neq0,1$}.
    \end{displaymath}
  \end{enumerate}
\end{theorem}
\section{Cohomology of full extended complex}
\label{s:full}
\renewcommand{\theequation}{\thesection .\arabic{equation}}
\setcounter{equation}{0}
In this section we will calculate the semi-relative cohomology of the
full extended complex. We will use the same technique as for the
chiral complex and obtain similar results.
\subsection{Review of semi-relative BRST cohomology}
\label{sec:srbrst}
\renewcommand{\theequation}{\thesubsection.\arabic{equation}}
By definition, the semi-relative cohomology $\H_S$ consists of $Q$
invariant states annihilated by $b_0-\ov b_0$, modulo $Q\ket{\Lm}$
where $\ket{\Lm}$ is annihilated by $b_0-\ov b_0$. For future
convenience we set $b_0^\pm = b_0\pm\ov b_0$ and $c_0^\pm =
(1/2)(c_0\pm\ov c_0)$.

Semi-relative cohomology $\H_S$ of the BRST complex can be easily
expressed in terms of relative cohomology $\H_R$. The latter consists
of the $Q$ invariant states annihilated both by $b_0$ and $\ov b_0$
modulo $Q\ket{\Lm}$ where $\ket{\Lm}$ is also annihilated by $b_0$ and
$\ov b_0$ separately. We can express $\H_S$ in terms of $\H_R$
using the following exact sequence (see ref.~\cite{wz92}):
\begin{equation}
  \label{exseq}
\cdots \to{\cal H}^n_R \onarrow{i}
{\cal H}_S^n \buildrel b_0^+ \over \longrightarrow
{\cal H}^{n-1}_R
\buildrel \{ Q , c_0^+ \} \over \longrightarrow
{\cal H}^{n+1}_R \onarrow{i} {\cal H}_S^{n+1}\to\cdots.
\end{equation}
The map $i$ is induced in the cohomology by the natural embedding of
the relative complex into the absolute complex. The relative
cohomology can be found as a tensor product of the relative
cohomologies of the left and right chiral sectors.

This is particularly easy for the case $p\neq0$ because in this case
the chiral relative cohomology is non-zero only at ghost number one
(see \cite{fgz86}) and, therefore, $\H_R$ consists of ghost number two
states only. The exact sequence (\ref{exseq}) reduces in this case to
\begin{equation}
  \label{redseq}
  0\to\H_S^3 \onarrow{b_0^+}\H_R^2\to0\qquad\mbox{and}\qquad
  0\to\H_R^2 \onarrow{i}\H_S^2\to0,
\end{equation}
which means that $\H_S^3$ is isomorphic to $\H_S^2$ which in turn is
isomorphic to $\H_R^2$ and there are no semi-relative cohomology
states at ghost numbers other than two or three. We can even write
explicit representatives in terms of dimension $(1,1)$ primary matter
states. Let $\ket{v,p}$ denote such a state. Representatives of
semi-relative cohomology classes can be written as
\begin{equation}
  \label{sreps}
  c_1\ov c_1\ket{v,p}\qquad\mbox{and}\qquad
  c_0^+c_1\ov c_1\ket{v,p}.
\end{equation}

\subsection{Semi-relative cohomology of $p\not\equiv0$ extended complex}
\label{s:nzclo}
\renewcommand{\theequation}{\thesubsection.\arabic{equation}} The
calculation of the extended cohomology for the semi-relative complex
in the case of $p\not\equiv0$ is line by line parallel to that of the
chiral one. When we add polynomials of $\px=(p\cdot x)$ the resulting
cohomology is given by ghost number two BRST states only, and the
semi-relative cohomology of the full extended complex is obtained by
adding polynomials of transverse components of $x$:
\begin{equation}
  \label{massemi}
  \Gr\,\H_S(\hQ,\hV_p)=\IC[\tx^1,\cdots,\tx^{D-1}]\otimes
        \H^2_S.
\end{equation}
The case of the massless ($p^2=0$) complex may require some special
consideration because there is no straightforward way to choose the
transverse variables $\tx^i$ but the answer is the same.

In our opinion, an important result is that the extended cohomology
does not contain ghost number three states. Let us explain in more
details what happens to the ghost number three semi-relative states
when we extend the complex by $x_0$. According to the results on
extended cohomology for every state $c^+_0c_1\ov c_1\ket{v,p}$ we
should be able to find a vector $\ket{w,p}$ in the extended complex
such that $c^+_0c_1\ov c_1\ket{v,p}=\hQ\ket{w,p}$. We have found the
leading part of $\ket{w,p}$ when we calculated the spectral sequence.
It is given by\footnote{This is not applicable to the massless case
  $p^2=0$ which has to be analyzed separately.} ${(p\cdot x)\over
  p^2}c_1\ov c_1\ket{v,p}$. Applying $\hQ$ to this state we get
\begin{equation}
  \label{almost}
  \hQ {(p\cdot x)\over p^2}c_1\ov c_1\ket{v,p}=
  c^+_0c_1\ov c_1\ket{v,p} +
  \sum_n p_\mu  c_1\ov c_1(c_{-n}\a_n^\mu+\ov
  c_{-n}\ov\a_n^\mu)\ket{v,p}.
\end{equation}
Thus in order to prove that $c^+_0c_1\ov c_1\ket{v,p}$ is trivial we
have to prove that the sum in the right hand side represents a $\hQ$
exact state. The latter is a direct consequence of the absence of
relative cohomology of ghost number three. Indeed, this sum is
annihilated by $b_0$, $\ov b_0$, and $Q$ and if it were nontrivial it
would have represented a non-zero cohomology class of ghost number
three in the relative complex.

\subsection{Semi-relative cohomology of $p=0$ extended complex}
\label{s:zclo}
\renewcommand{\theequation}{\thesubsection.\arabic{equation}}
\setcounter{equation}{0} The calculation of the cohomology of the
zero-momentum extended complex is based on the same ideas that we used
for the chiral case but is technically more difficult. The major
complication is due to a much larger BRST cohomology.

We can find the BRST cohomology of the zero momentum semi-relative complex
using the long exact sequence we mentioned above (see \eq{exseq})
which in this case breaks in to the following exact sequences:
\begin{equation}
  \label{shseqs}
  \begin{array}{rcl}
    0\to\H_R^0\onarrow{i}&\H_S^0&\onarrow{b_0^+}0,\\
    0\to
    \H_R^1\onarrow{i}&\H_S^2&  \onarrow{b_0^+}\H_R^0
    \onarrow{\{Q,c_0^+\}}\H_R^2,\\
    0\to
    \H_R^1\onarrow{i}&\H_S^3&  \onarrow{b_0^+}\H_R^0
    \onarrow{\{Q,c_0^+\}}\H_R^2,\\
    \H_R^2\onarrow{\{Q,c_0^+\}}
    \H_R^4\onarrow{i}&\H_S^4&  \onarrow{b_0^+}\H_R^3\to0,\\
    0\to&\H_S^5&  \onarrow{b_0^+}\H_R^4\to0.
  \end{array}
\end{equation}
The cohomology of the relative complex can be found as a tensor
product of the left and right relative cohomologies. The later can be
found, for example, in ref.~\cite{fgz86} and consists of ghost number
zero, one, and two states. Therefore, the relative complex has
non-trivial cohomologies at ghost numbers from zero through four and
their representatives and dimensions are listed in \tab{rcoh}.
\begin{table}[tbp]
  \begin{center}
    \leavevmode
    \begin{tabular}[c]{|c|c|c|}
      \hline
      Ghost \#&Representatives&Dimension\\ \hline\hline
      0&$\ket0$&$1$\\ \hline
      1&$c_1\a_{-1}^\mu\ket0,\quad\ov c_1\ov\a_{-1}^\mu\ket0$&$2D$\\ \hline
      2&$c_1\a_{-1}^\mu\ov c_1\ov\a_{-1}^\nu\ket0,\quad
         c_1c_{-1}\ket0,\quad\ov c_1\ov c_{-1})\ket0$&$D^2+2$\\ \hline
      3&$c_1\a_{-1}^\mu\ov c_1\ov c_{-1}\ket0,\quad
         c_1c_{-1}\ov c_1\ov\a_{-1}^\nu\ket0$&$2D$\\ \hline
      4&$c_1c_{-1}\ov c_1\ov c_{-1}\ket0$&$1$\\ \hline
    \end{tabular}
  \end{center}
  \caption{Cohomology of the relative BRST complex at $p=0$}
  \label{tab:rcoh}
\end{table}

Using the exact sequences (\ref{shseqs}) we can easily find the
dimensions and representatives of the semi-relative cohomology as
shown in \tab{scoh}.
\begin{table}[t]
  \begin{center}
    \leavevmode
    \begin{tabular}[c]{|c|c|c|}
      \hline
      Ghost \#&Representatives&Dimension\\ \hline\hline
      0&$\ket0$&$1$\\ \hline
      1&$c_1\a_{-1}^\mu\ket0,\quad\ov c_1\ov\a_{-1}^\mu\ket0$&$2D$\\ \hline
      2&$c_1\a_{-1}^\mu\ov c_1\ov\a_{-1}^\nu\ket0,\quad
         (c_1c_{-1}-\ov c_1\ov c_{-1})\ket0$&$D^2+1$\\ \hline
      3&$c_0^+c_1\a_{-1}^\mu\ov c_1\ov\a_{-1}^\nu\ket0,\quad
         c_0^+(c_1c_{-1}-\ov c_1\ov c_{-1})\ket0$&$D^2+1$\\ \hline
      4&$c_0^+c_1\a_{-1}^\mu\ov c_1\ov c_{-1}\ket0,\quad
          c_0^+c_1c_{-1}\ov c_1\ov\a_{-1}^\mu\ket0$&$2D$\\ \hline
      5&$c_0^+c_1c_{-1}\ov c_1\ov c_{-1}\ket0$&$1$\\ \hline
    \end{tabular}
  \end{center}
  \caption{BRST cohomology of the semi-relative complex at $p=0$}
  \label{tab:scoh}
\end{table}

The extended module $\hV_0$ is given by
\begin{equation}
  \label{chv0}
  \hV_0=\IC[x^0, \dots,x^{{D-1}}]\otimes V_0,
\end{equation}
and the operator $\hQ$ has the following $x$-degree decomposition
\begin{equation}
  \label{cddn}
  \hQ=\p_0+\p_1+\p_2,
\end{equation}
where
\begin{equation}
  \label{cdxpr}
  \begin{array}{rcl}
    \p_0&=&1\otimes Q,\\
    \p_1&=&\ds-i{\p\over\p x^\mu}\otimes
                    \sum_n(c_n\a_{-n}^\mu+\ov c_n\ov\a_{-n}^\mu),\\
    \p_2&=&\ds-{1\over2}\Box\otimes (c_0+\ov c_0)=-\Box\otimes c^+.\\
  \end{array}
\end{equation}
As it was the case for the chiral complex, the cohomology of $\p_0$
coincides with the cohomology of the semirelative BRST module, $\H_S$
tensored with $\IC[x^0,\dots,x^{{D-1}}]$ and the first term in the
spectral sequence is given by
\begin{equation}
  \label{cle1}
    E_1=H_S(\p_0,\hV_0)=\IC[x^0,\dots,x^{{D-1}}]\otimes \H_S
\end{equation}

Using information from Table~\ref{tab:scoh}, we can parameterize the
space $E_1=\IC[x^0\cdots x^{D-1}]\otimes\H_S$ as follows
\begin{eqnarray}
\nonumber
  (R^{[0]})&=&R^{[0]}\otimes\ket0,\\  \nonumber
  (P^{[1]}_\mu,\;\ov P^{[1]}_\mu)&=&P^{[1]}_\mu\otimes
  c_1\a_{-1}^\mu\ket0+
       \ov P^{[1]}_\mu\otimes\ov c_1\ov\a_{-1}^\mu\ket0,\\  \nonumber
  (Q^{[2]}_{\mu\nu},\; R^{[2]})&=&
   Q^{[2]}_{\mu\nu}\otimes c_1\a_{-1}^\mu\ov c_1\ov\a_{-1}^\nu\ket0+
         R^{[2]}\otimes (c_1c_{-1}-\ov c_1\ov c_{-1})\ket0,\\  \nonumber
  (Q^{[3]}_{\mu\nu},\; R^{[3]})&=&
   Q^{[3]}_{\mu\nu}\otimes c_0^+c_1\a_{-1}^\mu\ov c_1\ov\a_{-1}^\nu\ket0+
         R^{[3]}\otimes c_0^+(c_1c_{-1}-\ov c_1\ov c_{-1})\ket0,\\  \nonumber
  (P^{[4]}_\mu,\;\ov P^{[4]}_\mu)&=&P^{[4]}_\mu\otimes
  c_0^+c_1\a_{-1}^\mu\ov c_1\ov c_{-1}\ket0+
       \ov P^{[4]}_\mu\otimes c_0^+c_1c_{-1}\ov c_1\ov\a_{-1}^\mu\ket0,\\
  (R^{[5]})&=&R^{[0]}\otimes c_0^+c_1c_{-1}\ov c_1\ov c_{-1}\ket0,
  \label{e1reps}
\end{eqnarray}
where $Q^{\mu\nu}$, $P_\mu$ and $\ov P_\mu$ and $R$ are correspondingly
tensor-, vector-, and scalar-valued polynomials of $D$ variables
$x^\mu$. Applying $\p_1$ to the representatives given by
(\ref{e1reps}) and dropping $Q$-trivial states we can find that the first
differential acts according to the following diagram
\begin{equation}
  \label{d1acts}
  \begin{array}{rrcl}
0:&               (R^{[0]})      &\dn&(0)  \\
1:&     (P_\mu^{[1]},\; \ov P_\mu^{[1]})  &\dn&
         (\p_\mu R^{[0]},\; \p_\mu R^{[0]})\\
2:&  ( Q_{\mu\nu}^{[2]},\;R^{[2]})
                        &\dn&
                ( \p_\mu\ov P_\nu^{[1]} - \p_\nu P_\mu^{[1]},\;
                        \p^\mu\ov P_\mu^{[1]} - \p^\mu P_\mu^{[1]})\\
3:&  ( Q_{\mu\nu}^{[3]},\;R^{[3]})
                        &\dn&(0,\; 0)\\
4:&     (P_\mu^{[4]},\; \ov P_\mu^{[4]})  &\dn&
(\p^\nu Q_{\mu\nu}^{[3]}+\p_\mu R^{[3]},
                           \;\p^\nu Q_{\nu\mu}^{[3]}+\p_\mu R^{[3]})\\
5:&               (R^{[5]})      && (\p_\mu P^{[4]}-\p_\mu\ov P^{[4]})
  \end{array}
\end{equation}
Consider the cohomology of $d_1$. From the explicit formulae
(\ref{d1acts}) it is clear that $E_2=H(d_1,E_1)$ contains no ghost
number five states and the only state that survives at ghost number
zero is given by a constant $R^{[0]}$ and is the vacuum. It is less
trivial to show that $E_2^{(4)}=0$, and that $E_2^{(1)}$ contains only
$x$-degree one and two states. We will prove these two results in
\ref{app:l} (lemmas~\ref{l:h4} and~\ref{l:h1}). This shows that the
space $E_2$ looks as shown in \fig{zclo} (left).
\setlength{\unitlength}{20pt}%
\savebox{\dddt}{%%
  \begin{picture}(1,0)%%
    \multiput(0,0)(0.1,0){11}{\dt}%
  \end{picture}%
  }%
\savebox{\ddt}{%%
  \begin{picture}(1,0)%%
    \multiput(0.05,0)(0.1,0){10}{\dt}%
  \end{picture}%
  }%
\savebox{\shade}{%%
  \begin{picture}(1,1)%
    \multiput(0,0)(0,0.1){11}{\usebox{\dddt}}%
    \multiput(0,0.05)(0,0.1){10}{\usebox{\ddt}}%
  \end{picture}%
}%
\savebox{\halfshade}{%%
  \begin{picture}(1,1)%
    \multiput(0,0)(0,0.1){5}{\usebox{\dddt}}%
    \multiput(0,0.05)(0,0.1){5}{\usebox{\ddt}}%
  \end{picture}%
}%
\savebox{\sgrid}{%%
  \begin{picture}(6,6)%
    \multiput(0,-0.5)(1,0){6}{\line(0,1){6}}%
    \multiput(-0.5,0)(0,1){6}{\line(1,0){5.5}}%
    \put(5.05,-0.5){\line(0,1){6}}%
    \setcounter{nmb}{-1}%
    \multiput(-0.7, 0)(0,1){5}{\addtocounter{nmb}{1}%%
      \makebox(0,1){\small$\arabic{nmb}$}}%
    \put(-0.7,5){\vector(0,1){0.5}}%
    \put(-0.7,5.6){\makebox(0,0)[b]{$s$}}%
    \setcounter{nmb}{-5}%
    \multiput(0, -0.7)(1,0){5}{\addtocounter{nmb}{1}%%
      \makebox(1,0){\small$\arabic{nmb}$}}%
    \put(5,-0.7){\vector(1,0){0.5}}%
    \put(5.6,-0.7){\makebox(0,0)[l]{$r$}}%
  \end{picture}%
  }%
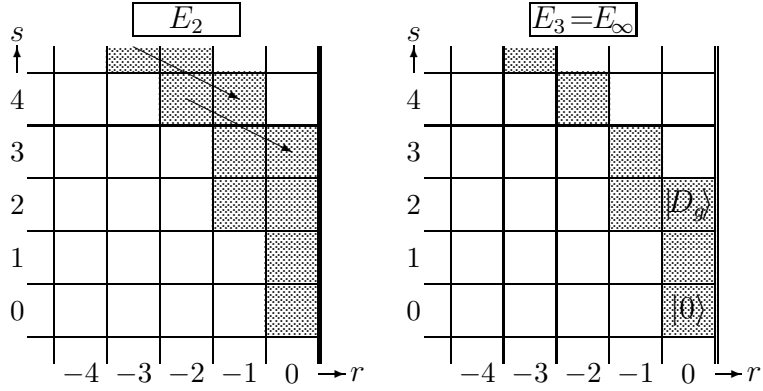
\begin{figure}[tbp]%
  \begin{center}%
    \leavevmode%
    \begin{picture}(5.5,6.5)(-1,-1)%
      \ifpic%
      \put(0,0){\usebox{\sgrid}}%
      \multiput(1,5)(1,0){2}{\usebox{\halfshade}}%
      \multiput(4,3)(-1,1){2}{\usebox{\shade}}%
      \multiput(4,2)(-1,1){3}{\usebox{\shade}}%
      \multiput(4,1)(-1,1){2}{\usebox{\shade}}%
      \put(4,0){\usebox{\shade}}%
      \multiput(2.5,4.5)(-1,1){2}{\vector(2,-1){2}}%
      \put(1.5,5.7){\framebox(2,.6){{$E_2$}}}%
      \fi%
    \end{picture}\hspace{2\unitlength}%
    \begin{picture}(5.5,6.5)(-1,-1)%
      \ifpic%
      \put(0,0){\usebox{\sgrid}}%
      \multiput(1,5)(1,0){1}{\usebox{\halfshade}}%
      \multiput(4,2)(-1,1){3}{\usebox{\shade}}%
      \multiput(4,1)(-1,1){2}{\usebox{\shade}}%
      \put(4,0){\usebox{\shade}}%
      \put(4,0){\makebox(1,1){\bf$\ket0$}}%
      \put(4,2){\makebox(1,1){\bf$\ket{\!D_g\!}$}}%
      \put(1.5,5.7){\framebox(2,.6){{$E_3\!=\!\!E_{\!\infty}$}}}%
      \fi%
    \end{picture}%
  \end{center}%
  \caption{Spectral sequence for the zero momentum closed string states.}%
  \label{fig:zclo}%
\end{figure}%
{}From the structure of $E_2$ we conclude that $d_2$ may only act from
$E_2^{(2)}$ to $E_2^{(3)}$. In order to find
$$
d_2\;(Q_{\mu\nu}^{[2]},\;R^{[2]})=
(\p_2-\p_1\p_0^{-1}\p_1)\;(Q_{\mu\nu}^{[2]},\;R^{[2]}),
$$
we have to keep the $\p_0$ trivial terms when we apply $\p_1$ to $(
Q_{\mu\nu}^{[2]},\;R^{[2]})$ which were dropped in \eq{d1acts}. Using
(\ref{e1reps}) and (\ref{cdxpr}) we infer
\begin{equation}
  \label{p1ful}
  \begin{array}{r}
  \p_1\;(Q_{\mu\nu}^{[2]},\;R^{[2]})=
      \p_0\left((\p^\nu Q_{\mu\nu}^{[2]}+\p_\mu R^{[2]})\otimes
           c_0^+\ov c_1\ov\a_{-1}^\mu\ket0\right.\\ \left.
               -(\p^\mu Q_{\mu\nu}^{[2]}+\p_\nu R^{[2]})\otimes
           c_0^+c_1\a_{-1}^\nu\ket0\right).
  \end{array}
\end{equation}
Now we can apply $\p_1$ to the argument of $\p_0$ above and add the
image of $\p_2$ to obtain
\begin{equation}
  \label{d2c}
  \qquad d_2 ( Q_{\mu\nu}^{[2]},\;R^{[2]})=( \dl Q_{\mu\nu}^{[3]},\;\dl
  R^{[3]}),
\end{equation}
where
\begin{equation}
\begin{array}{l}
\dl Q_{\mu\nu}^{[3]}=\Box Q_{\mu\nu}^{[2]}-
     \p^\lm\p_\mu Q_{\lm\nu}^{[2]}-\p^\lm\p_\nu Q_{\mu\lm}^{[2]}+
      2\,\p_\mu\p_\nu R^{[2]},\\
\dl R^{[3]}=-\p^\lm\p^\rho Q_{\lm\rho}^{[2]}+
      2\,\Box R^{[2]}.
\end{array}
\end{equation}
As we will prove in \ref{app:l}, this map is surjective and therefore
the cohomology at ghost number three is trivial. The structure of
$E_3$ is presented in \fig{zclo} (right). It is obvious that the
spectral sequence collapses at $E_3$ and $E_\infty=E_3$. Therefore, we
obtain the result that is similar to that in the chiral case (see
\eq{p0answer}):
\begin{equation}
  \label{cloansw}
\Gr_rH^{r+s}(\hQ,\hV_0)=E^{r,s}_\infty=E_3^{r,s}.
\end{equation}
Note that $d_2$ acts non-trivially only between ghost number two and ghost
number three states, at the point where $d_1$ vanishes. This
observation allows us to combine the calculation of $E_2$ and $E_3$
into one cohomology problem for the following complex
\begin{equation}
  \label{vcmpl}
  0\to
  V^{(0)}\onto{d_1}V^{(1)}\onto{d_1}V^{(2)}
  \onto{d_2}V^{(3)}\onto{d_1}V^{(4)}
  \onto{d_1}V^{(5)}\to0,
\end{equation}
where $V^{(k)}=\bigoplus_r E_1^{r,k-r}$ are the ghost number $k$
subspaces of $E_1$. This is very similar to what we did for the chiral
complex (\eq{e3df} and \eq{2in1}), but, unfortunately, we lack a
simple geometrical interpretation for the complex above.

Since $d_1$ and $d_2$ have $x$-degree $-1$ and $-2$ correspondingly,
we can decompose the complex above into the sum of the following
subcomplexes
\begin{equation}
  \label{ncmpl}
  0\to
  V^{(0)}_{n+2}\onto{d_1}
  V^{(1)}_{n+1}\onto{d_1}
  V^{(2)}_{n}\onto{d_2}
  V^{(3)}_{n-2}\onto{d_1}
  V^{(4)}_{n-3}\onto{d_1}
  V^{(5)}_{n-4}\to0,
\end{equation}
where the subscripts refer to $x$-degree.  Using this decomposition
and the results of \ref{app:l} one can easily calculate the dimensions
of $\Gr_rH_S$ spaces.

Let us summarize our results on the semi-relative cohomology of the
extended complex.
\begin{theorem}
  The semi-relative cohomology of the extended complex $\hV_p$, admits
  a natural filtration (by the {\em minus\/} $x$-degree) $\F_r
  H_S(\hV_p,\hQ)$.  The cohomology can be described using the
  associated graded spaces
  $$\Gr_r H_S(\hV_p,\hQ)=\F_r H_S(\hV_p,\hQ)/\F_{r+1} H_S(\hV_p,\hQ)$$
  as follows
  \begin{enumerate}
  \item Non-zero momentum ($p\not\equiv0$): cohomology is trivial unless
    $p^2=2n-2$, where $n$ is a non-negative integer. In the latter
    case
    \begin{eqnarray*}
      \dim \Gr_r H^2_S(\hV_p,\hQ)&=&\ds{24-r\choose 24}\,d_n^2,\\
      \dim H^l_S(\hV_p,\hQ)&=&0\qquad\mbox{for $l\neq2$}
    \end{eqnarray*}
    where $d_n$  are generated by the following partition function
    $$\prod_{k=1}^\infty(1-z^k)^{-24}=\sum_{n=1}^\infty d_n z^n.$$
    see ref.~\cite{GSchW}.
  \item Zero momentum case ($p\equiv0$): the cohomology appears at
    ghost number zero (the vacuum)
    \begin{displaymath}
            \dim H^0_S(\hV_p,\hQ)=1,
    \end{displaymath}
    ghost number one (global symmetries of the background, the
    Poincar\'e algebra),
    \begin{displaymath}
      \dim \Gr_r H^1_S(\hV_p,\hQ)=\ds\cases{
        \hfil{D}\hfil& if $r=0$;\cr\cr
        \ds{D(D-1)\over2}& if $r=-1$;\cr\cr
        \hfil0\hfil&otherwise;\cr}
    \end{displaymath}
    and ghost number two (physical states),
    \begin{eqnarray*}
      \dim \Gr_{0} H^2_S(\hV_p,\hQ)&=&1,\\
      \dim \Gr_{-1} H^2_S(\hV_p,\hQ)&=&\ds{D\,(D^2-3\,D+8)\over6},\\
      \dim \Gr_{-2} H^2_S(\hV_p,\hQ)&=&{D(D-2)(D+2)\over3} ,\\
      \dim \Gr_{-3} H^2_S(\hV_p,\hQ)&=&\ds{{( D + { 2} ) \,
          (5\,D^3-16\,D^2+15\,D-12)}
        \over {24}},\\
      \dim \Gr_{r} H^2_S(\hV_p,\hQ)&=&\chi_r\mbox{ for $r\leq-4$.}
    \end{eqnarray*}
    where
    \begin{displaymath}
      \begin{array}{rcl}
      \chi_r&=&\ds(D^2+1)\left[{D-1-r\choose D-1}-{D-3-r\choose
          D-1}\right]\\[12pt]
      &&\hfill+\ds
      2D\left[{D-r-4\choose D-1}- {D-r\choose D-1}\right]\\[12pt]
      &&\hfill+\ds{D-r+1\choose D-1}-{D-r-5\choose D-1}.
      \end{array}
    \end{displaymath}
    There are no non-trivial cohomology states at any other ghost
    number,
    \begin{displaymath}
      \dim H^l_S(\hV_p,\hQ)=0\qquad\mbox{for $l\neq0,1,2$.}
    \end{displaymath}
  \end{enumerate}
\end{theorem}

There is no evident structure among the ghost number two cohomology
states. In the next section we are going to make some order in this
zoo of zero momentum physical states using their transformation
properties under the Lorentz group.
\section{Lorentz group and extended complex}
\label{s:lor}
Since the cohomology of the extended complex is designed to describe
the string theory near the flat background, the Lorentz group should
act on it naturally, mapping the cohomology states to the cohomology
states.  In this section we investigate the action of the generators
$J^{\mu\nu}$ of the infinitesimal Lorentz transformations.
\subsection{Lorentz group acting on non-zero momentum extended complex}
\label{s:lornz}
\renewcommand{\theequation}{\thesubsection.\arabic{equation}}
\setcounter{equation}{0} The generators of the infinitesimal Lorentz
transformations, or the angular momentum operators of the closed
string can be written as \cite{GSchW}
\begin{equation}
  \label{jdef}
  J^{\mu\nu}=l^{\mu\nu}+E^{\mu\nu}+\ov{E}^{\mu\nu},
\end{equation}
where
\begin{equation}
  \begin{array}{rcl}
  l^{\mu\nu}&=&x^\mu_0\a_0^\nu-x^\nu_0\a_0^\mu,\\
  E^{\mu\nu}&=&\ds-i\sum_{n=1}^\infty{1\over n}\,
            (\a_{-n}^\mu\a_n^\nu-\a_{-n}^\nu\a_n^\mu),\\
  \ov{E}^{\mu\nu}&=&\ds -i\sum_{n=1}^\infty{1\over n}\,
            (\ov\a_{-n}^\mu\ov\a_n^\nu-\ov\a_{-n}^\nu\ov\a_n^\mu).
  \end{array}
\end{equation}
Applying the zero mode part $l^{\mu\nu}$ to $\hV_p$, we obtain
\begin{equation}
  \label{zml}
  l^{\mu\nu}=x^\mu\otimes p^\nu-x^\nu\otimes p^\mu+
             x^\mu{\p\over\p x_\nu}\otimes1-
             x^\nu{\p\over\p x_\mu}\otimes1.
\end{equation}
One can easily check that the $J^{\mu\nu}$ operators commute with
$\hQ$ and therefore are well defined on the cohomology. Note that for
$p\not\equiv0$ these operators mix states of different $x$-degree.

Consider the massive case, $p^2\not\equiv0$. Recall that any physical state
in the extended complex can be represented by
\begin{equation}
  \label{phys2}
  P(\tx^1,\cdots,\tx^{D-1})\otimes c_1\ov c_1\ket{v,p}+\cdots\,,
\end{equation}
where $\ket{v,p}$ is a dimension $(1,1)$ primary matter state with the
momentum $p$ and dots stand for the lower $x$-degree terms. Suppose
the transverse components are chosen as
\begin{equation}
  \label{transv}
  \tx^i=x^i-{p^i(p\cdot x)\over p^2}.
\end{equation}
We can obtain a state with the same leading term by applying the
following operator to $c_1\ov c_1\ket{v,p}$
\begin{equation}
  \label{appl}
  P\left({p_\mu J^{1\mu}\over p^2},\cdots,
         {p_\mu J^{D-1\mu}\over p^2}\right)\; c_1\ov c_1\ket{v,p}=
   P(\tx^1,\cdots,\tx^{D-1})\otimes c_1\ov c_1\ket{v,p}+\cdots       \;.
\end{equation}
Together with the results on the cohomology of the extended complex
obtained in the previous section, \eq{massemi}, this shows that the
semi-relative cohomology of a $p^2\neq0$ extended complex is spanned
by the states which are generated from the standard physical states by
the infinitesimal Lorentz boosts. Although the analysis above fails
for the massless states $p^2=0$ ($p\not\equiv0$), the result is still
the same. One can check it using the explicit formulae for the
cohomology states at $p^2=0$.

We conclude that the ghost number two cohomology of the extended
complex at $p\not\equiv0$ has the same physical contents as that of
the BRST complex.

\subsection{Lorentz group and the zero momentum states}
\label{s:lorz}
\renewcommand{\theequation}{\thesubsection.\arabic{equation}}
\setcounter{equation}{0}
The Lorentz generators, $J^{\mu\nu}$, act on the zero-momentum
states as linear operators of zero $x$-degree:
\begin{equation}
  \label{zjmn}
  J^{\mu\nu}= x^\mu{\p\over\p x_\nu}\otimes1-
                x^\nu{\p\over\p x_\mu}\otimes1+
                1\otimes (E^{\mu\nu}+
                \ov E^{\mu\nu}).
\end{equation}
\arraycolsep 0pt
\begin{table}[t]
  \begin{center}
    \leavevmode
    \small
    \hbox to \textwidth{\hss
    \begin{tabular}{|c|c|c|c|c|c|}
      \hline
      $V_{n+2}^{(0)}$&$V_{n+1}^{(1)}$&$V_{n}^{(2)}$&
      $V_{n-2}^{(3)}$&$V_{n-3}^{(4)}$&$V_{n-4}^{(5)}$\\
      \hline
      \hline
      $\rH_{n+2}$&$2\,\rH_{n+2}         $&$\rH_{n+2}$&&&\\
      \hline
      $\rH_{n}  $&$4\,\rH_{n}\!+\!2\,\rV_{n}    $&$
      \begin{array}{l}
        5\,\rH_{n}\!+\!\rS_{n}\\
        \!+\!\rA_{n}\!+\!2\,\rV_{n}
      \end{array}
      $&
      $\rH_{n}$&&\\
      \hline
      $\rH_{n-2}$  &   $4\,\rH_{n-2}\!+\!2\,\rV_{n-2}$ &
      $\begin{array}{l}
      6\,\rH_{n-2}\!+\!\rS_{n-2}\\
      \!+\!\rA_{n-2}\!+\!4\,\rV_{n-2}
      \end{array}$
      &
      $\begin{array}{l}
        5\,\rH_{n-2}\!+\!\rS_{n-2}\\
        \!+\!\rA_{n-2}\!+\!2\,\rV_{n-2}
      \end{array}$
      &$2\,\rH_{n-2}$&\\
      \hline
      $\rH_{n-4}$&$4\,\rH_{n-4}\!+\!2\,\rV_{n-4}$&
      $\begin{array}{l}
        6\,\rH_{n-4}\!+\!\rS_{n-4}\\
        \!+\!\rA_{n-4}\!+\!4\,\rV_{n-4}
      \end{array}$
      &
      $\begin{array}{l}
        6\,\rH_{n-4}\!+\!\rS_{n-4}\\
        \!+\!\rA_{n-4}\!+\!4\,\rV_{n-4}
      \end{array}$
      &$2\,\rH_{n-4}\!+\!2\,\rV_{n-4}$&$\rH_{n-4}$\\
      \hline
      $\rH_{n-6}$&$4\,\rH_{n-6}\!+\!2\,\rV_{n-6}$&
      $\begin{array}{l}
        6\,\rH_{n-6}\!+\!\rS_{n-6}\\
        \!+\!\rA_{n-6}\!+\!4\,\rV_{n-6}
      \end{array}$
      &
      $\begin{array}{l}
        6\,\rH_{n-6}\!+\!\rS_{n-6}\\
        \!+\!\rA_{n-6}\!+\!4\,\rV_{n-6}
      \end{array}$
      &$2\,\rH_{n-6}\!+\!2\,\rV_{n-6}$&$\rH_{n-6}$\\
      \hline
      $\cdots$&$\cdots$&$\cdots$&$\cdots$&$\cdots$&$\cdots$\\
      \hline
    \end{tabular}\hss}
  \end{center}
  \caption{Decomposition of the complex (\protect\ref{cpl}) into irreducible
    representations of $SO(D-1,1)$ for $n\geq2$.}
  \label{tab:irreps}
\end{table}

Consider the complexes
\begin{equation}
  \label{cpl}
  0\to
  V^{(0)}_{n+2}\onto{d_1}
  V^{(1)}_{n+1}\onto{d_1}
  V^{(2)}_{n}\onto{d_2}
  V^{(3)}_{n-2}\onto{d_1}
  V^{(4)}_{n-3}\onto{d_1}
  V^{(5)}_{n-4}\to0.
\end{equation}
which calculate $\Gr H_S(\hV_0,\hQ)$. By definition $V^{(0)}_n$ and
$V^{(5)}_n$ are the spaces of homogeneous polynomials of degree $n$.
These spaces are reducible under the Lorentz group because the
subspaces of the polynomials of the form $(x^\mu x_\mu)^{k} h_{n-2k}$
are invariant under $SO(D-1,1)$. Furthermore, if $h_{n-2k}$ are
harmonic, $\Box h_{n-2k}=0$, these subspaces form irreducible
representations of $SO(D-1,1)$. We will denote these irreducible
representations by $\rH_n$.  These representations can be
alternatively described by Young tableaux as
\setlength{\unitlength}{10pt}
\begin{equation}
  \label{hyoung}
  \rH_n=
  \overbrace{%
    \begin{picture}(5,1)%
      \put(0,0){\line(1,0){5}}%
      \put(0,1){\line(1,0){5}}%
      \multiput(0,0)(1,0){3}{\line(0,1){1}}%
      \multiput(4,0)(1,0){2}{\line(0,1){1}}%
      \put(2,0){\makebox(2,1){$\cdots$}}%
    \end{picture}}^n\;.
\end{equation}
Now we can write the decomposition of $V^{(0)}_n$ or $V^{(5)}_n$ into
irreducible representations as
\begin{equation}
  \label{05irreps}
  V^{(0)}_n=V^{(5)}_n= \rH_n+\rH_{n-2}+\rH_{n-4}+\cdots\;.
\end{equation}
At the ghost numbers one and four spaces we find another kind of
irreducible representations:
\begin{equation}
  \label{vyoung}
  \rV_n=
  \overbrace{%
  \begin{picture}(5,1)%
    \put(0,0){\line(1,0){6}}%
    \put(0,1){\line(1,0){6}}%
    \put(0,-1){\line(1,0){1}}%
    \put(2,0){\line(0,1){1}}%
    \multiput(0,-1)(1,0){2}{\line(0,1){2}}%
    \multiput(4,0)(1,0){3}{\line(0,1){1}}%
    \put(2,0){\makebox(2,1){$\cdots$}}%
  \end{picture}}^n\kern\unitlength\;,
\end{equation}
and, finally in the decompositions of $V^{(2)}_n$ and $V^{(3)}_n$ we
will encounter
\begin{equation}
  \label{asyoung}
   \rA_n=
  \overbrace{
  \begin{picture}(5,1)
    \put(0,0){\line(1,0){5}}
    \put(0,1){\line(1,0){5}}
    \multiput(0,-1)(0,-1){2}{\line(1,0){1}}
    \put(2,0){\line(0,1){1}}
    \multiput(0,-2)(1,0){2}{\line(0,1){3}}
    \multiput(4,0)(1,0){2}{\line(0,1){1}}
    \put(2,0){\makebox(2,1){$\cdots$}}
  \end{picture}}^n\;\quad\vtop to 3\unitlength{\hbox{and}\vfill}\quad
 \rS_n=
  \overbrace{
  \begin{picture}(5,1)
    \put(0,0){\line(1,0){5}}
    \put(0,1){\line(1,0){5}}
    \put(0,-1){\line(1,0){2}}
    \put(2,0){\line(0,1){1}}
    \multiput(0,-1)(1,0){3}{\line(0,1){2}}
    \multiput(4,0)(1,0){2}{\line(0,1){1}}
    \put(2,0){\makebox(2,1){$\cdots$}}
  \end{picture}}^n\;.
\end{equation}

Suppose $n\geq2$.  \tab{irreps} shows the decomposition of the whole
complex~(\ref{cpl}) into irreducible representations.  From the series
of lemmas presented in \ref{app:l}, we know that the complex
(\ref{cpl}) has cohomology only in $V_n^{(2)}$. Using \tab{irreps} we
conclude that
\begin{equation}
  \label{cohrep}
  \Gr_{-n} H^2_S(\hV_0,\hQ)=\rH_n+\rA_n+\rS_n,
\end{equation}
which we obtain by ``subtracting'' the odd columns from column two and
``adding'' the even columns. A more detailed analysis shows that if
we choose the representatives of $\Gr_{-n} H^2_S(\hV_0,\hQ)$ so that
they belong to $\rH_n$, $\rA_n$, or $\rS_n$, they also will represent
cohomology classes of $\hQ$, no lower $x$-degree corrections required.

 Two exceptional cases,  $n=0$ and $n=1$,  have to be treated
separately. The decompositions of the complex~(\ref{cpl}) into irreducible
representations for these the first case is presented in \tab{irrep0}.
\begin{table}[ht]
  \begin{center}
    \leavevmode
    \begin{tabular}{|c|c|c|}
      \hline
      $V_{2}^{(0)}$&$V_{1}^{(1)}$&$V_{0}^{(2)}$\\
      \hline
      \hline
      $\rH_{2}$&$2\,\rH_{2}         $&$\rH_{2}$\\
      \hline
      $\rH_{0}  $&$2\,\rH_0+2\,\rV_{0}$&
      $\rH_{0}\!+\!\rV_{0}$\\
      \hline
    \end{tabular}\hfill
    \begin{tabular}{|c|c|c|}
      \hline
      $V_{3}^{(0)}$&$V_{2}^{(1)}$&$V_{1}^{(2)}$\\
      \hline
      \hline
      $\rH_{3}$    &$2\,\rH_{3}$ &  $\rH_{3}$\\
      \hline
      $\rH_{1}$    &$4\,\rH_{1}+2\,\rV_{1}$&
      $4\,\rH_{1}+\rA_1+2\,\rV_{1}$\\
      \hline
    \end{tabular}
    \caption{Decomposition of the complex (\protect\ref{cpl}) into
      irreducible representations for $n=0$ (left) and $n=1$ (right).}
    \label{tab:irrep0}
  \end{center}
\end{table}
Using the results of \hbox{\ref{app:l}} we can infer from \tab{irrep0}
(left) that
\begin{equation}
  \label{repgr0}
  \Gr_{-1} H_S^1(\hQ,\hV_0)=\rV_0\quad\mbox{and}\quad
  \Gr_0 H_S^2(\hQ,\hV_0)=\rH_0,
\end{equation}
and from \tab{irrep0} (right) that
\begin{equation}
  \label{repgr1}
  \Gr_{-1} H_S^2(\hQ,\hV_0)=\rH_1+\rA_1;
\end{equation}
and again, if we pick the representatives of $\Gr H_S$ from the
irreducible representations, they will be annihilated by $\hQ$ and
therefore represent the cohomology classes without lower $x$-degree
corrections.  For this case this can be easily checked by explicit
calculation (see \ref{app:l}).

It is tempting to interpret the irreducible
representations $\rH_n$, $\rS_n$, and $\rA_n$ as the dilaton,
graviton, and antisymmetric tensor. If we do so, it is not quite clear
why we have infinitely many irreducible representations for each
field, and not just one. We speculate that these representations are
related by infinitesimal shifts (the translational part of the
Poincar\'e algebra), which acts on the spaces of polynomials by
differentiation with respect to $x^\mu$.
\renewcommand{\thesection}{}
\section{Acknowledgments}
We would like to thank Barton Zwiebach for reading the manuscript and
giving many valuable suggestions. This work much benefited from
discussions with Alexander Gorohovsky and Barton Zwiebach. One of us
(A.B.) would also like to acknowledge conversations with Jeffrey
Goldstone, Kenneth Johnson and Paul Mende.
\newpage
\appendix
\renewcommand{\thesection}{Appendix~\Alph{section}}
\section{Spectral sequence}
\renewcommand{\thesection}{\Alph{section}}
\label{app:spectral}
\setcounter{equation}{0}
\renewcommand{\theequation}{\thesection.\arabic{equation}} In this
section we review some basic facts about a particular type of the
spectral sequence which we use in our analysis of the extended
complex. This is not intended to be a complete introduction to the
method. Our only goal is to introduce the spaces $E_0$, $E_1$ and
$E_2$ equipped with differentials $d_0$, $d_1$ and $d_2$ acting on
them. We will prove that these differentials have zero square and
provide some motivations to why their cohomologies are related to $\Gr
H$. For a more detailed analysis of the first three terms of a
spectral sequence, the reader is referred to the book by Dubrovin,
Fomenko and Novikov \cite{DNF3}.  A general introduction to the
spectral sequences from the physicist's point of view and further
references can be found in refs.~\cite{ffk89,dix91}.

Let $(C,d)$ be a complex with additional grading $C=\bigoplus C_r$
such that the differential $d$  can be written as
\begin{equation}
  \label{defdeg}
  d=\p_0+\p_1+\p_2,
\end{equation}
where $\p_n$ maps $C_r$ to $C_{r+n}$.  Since $d$ mixes vectors from
different gradings we can not define grading on cohomology $H(d, C)$,
but we can still define a decreasing filtration.  By filtration of the
element $x\in H(d,\Gr C)$ we will mean the smallest (negative) integer
$s$ such that $x$ is representable by a cocycle
\begin{equation}
  \label{cocycle}
\ov x =x_r +x_{r+1} +\cdots,
\end{equation}
where $x_r\in C_r$. We will denote the space of such vectors by
$F_rH(d,C)$. Using the filtration we can define a graded space
associated to the cohomology $H(d,C)$
\begin{equation}
  \label{grh}
   \Gr H=\bigoplus_s\Gr_s H,\quad\mbox{where}\quad\Gr_sH=F_sH/F_{s+1}H.
\end{equation}
The investigation of the spaces $\Gr_s H^n$ is carried out using the
method of ``successive approximations'' based on what is called the
``spectral sequence''. The idea is to construct a sequence of
complexes $(E_n,d_n)$ such that $E_{n+1}=H(d_n, E_n)$ which converges
to $\Gr H$
\begin{equation}
  \label{conv}
  \Gr_s H^n=E_{\infty}^{s,n-s}.
\end{equation}
Differentials $d_n$ are acting on the spaces $E_n^{r,s}$ as follows
\begin{equation}
  \label{dkact}
  d_n:\quad E_n^{r,s}\to E_n^{r+n, s-n+1}
\end{equation}
For a complete description of the spectral sequence and the proof of
the theorem which states that the spectral sequence converges to $\Gr
H$ we refer to \cite{DNF3,McC}. Let us describe the first few terms of
the spectral sequence.  Suppose $\ov x$ given in \eq{cocycle}
represent a cohomology class $x$ in $F_sH(d, C)$. Applying
$d=\p_0+\p_1+\p_2$ to $\ov x$ we obtain
\begin{equation}
  \label{dx}
  \begin{array}{l}
  d\ov x=\p_0 \ov x_r +(\p_1 \ov x_r + \p_0 \ov x_{r+1})+
   (\p_2 \ov x_r + \p_1 \ov x_{r+1}+\p_0 \ov x_{r+2})\\
   \kern2in+(\p_2 \ov x_{r+1} + \p_1 \ov x_{r+2}+\p_0 \ov x_{r+3})+\cdots.
  \end{array}
\end{equation}
where we enclosed in braces the terms from the same $C_r$ space. It
follows that
\begin{equation}
  \label{ppp}
\p_0\ov x_r=0,\quad \p_1\ov x_r=-\p_0 \ov x_{r+1},\quad
\p_2 \ov x_r=-\p_1\ov x_{r+1}-\p_0\ov x_{r+2}, \dots,
\end{equation}
form which we conclude that $\ov x_r$ is a $\p_0$ cocycle and a $\p_1$
cocycle modulo image of $\p_0$. This suggests that the first
approximation in the spectral sequence should be $E_1=H(\p_0, C)$ and
$d_0=\p_0$, the second approximation is $E_2=H(d_1, E_1)$, where $d_1$
is induced on $E_1=H(\p_0, C)$ by $\p_1$.  Using the second equation
of (\ref{ppp}) we can formally find $\ov x_{r+1}$ in terms of $\ov
x_r$ as $\ov x_{r+1}=-\p_0^{-1}\p_1\ov x_r$ and rewrite the last
equation of (\ref{ppp}) as
\begin{equation}
  \label{d2de}
  (\p_2 -\p_1\p_0^{-1}\p_1)\ov x_r=-\p_0\ov x_{r+2}.
\end{equation}
The last formulae suggests that $d_2$ is induced on $E_2=H(d_1, E_1)$
by $\p_2 -\p_1\p_0^{-1}\p_1$ and the third approximation is
$E_3=H(d_2, E_2)$.

Let us show that these differentials are well defined and square to
zero. For $d_0$ the first is obvious since it acts on the same space
as $\p_0$ and $\p_0^2=0$ follows from $d^2=0$ which is equivalent to
\begin{equation}
  \label{d2zero}
  \p_0^2=\{\p_0,\p_1\}=\p_1^2+\{\p_0,\p_2\}=\p_2^2=0.
\end{equation}

In order to show that $d_1$ is well defined we have to show that
$\p_1$ maps $\p_0$-closed vectors to $\p_0$-closed vectors and
$\p_0$-trivial to $\p_0$-trivial. This easily follows from the
anticommutation relation $\{\p_0,\p_1\}=0$. Let us prove that
$d_1^2=0$. Suppose $x\in E_1$ can be represented by a cocycle $\ov
x\in C$, $\p_0\ov x=0$. Then applying $\p^2_1=-\{\p_0,\p_2\}$ to $\ov
x$ we obtain a trivial cocycle $\p^2_1\ov x= -\p_0\p_2\ov x$.  In
cohomology this implies that $d_1^2 x=0$.

Before we consider the differential $d_2$, let us describe the space
$E_2$ on which it acts in greater details. By definition
$E_2=H(d_1,E_1)$, but $E_1$ in turn is the cohomology of the original
complex with respect to $\p_0$. Therefore, in order to find a $d_1$
cocycle we should start with a $\p_0$ cocycle $\ov x$ and require that
its image under $\p_1$ is $\p_0$ exact. Two cocycles $\ov x$ and $\ov
x'$ represent the same $d_1$ cohomology class if $\ov x'-\ov
x\in\im\p_0+\p_1\ker\p_0$. Let $\ov x$ be a $d_1$ cocycle, which means
that
\begin{equation}
  \label{d1cocy}
  \p_0\ov x=0\qquad \mbox{and}\qquad  \p_1\ov x=\p_0\ov y.
\end{equation}
We define $d_2\ov x$ as
\begin{equation}
  \label{d2def}
  d_2\ov x= \p_2\ov x - \p_1\ov y=(\p_2 -\p_1\p_0^{-1}\p_1)\ov x .
\end{equation}
Let us show that the result is again a $d_1$ cocycle. Indeed, using
the properties of $\p_n$ listed in \eq{d2zero} we obtain
$$
\p_0d_2\ov x= \p_0\p_2\ov x+\p_0\p_1\ov y=
\p_0\p_2\ov x-\p_1\p_0\ov y=\{\p_0,\p_2\}\ov x - \p_1^2\ov x=0
$$
and
$$
\p_1d_2\ov x= \p_1\p_2\ov x-\p_1^2\ov y=
\p_0\p_2\ov y.
$$

Similarly we can prove that if $\ov x$ and $\ov x'$ belong to the same
$d_1$ cohomology class then their $d_2$ images belong to the same
class as well. This will finally establish the correctness of the
definition of $d_2$ as an operator on $E_2$. In conclusion let us show
that $d_2^2=0$. With $\ov x$ as above we have
\[
\begin{array}{rcl}
d_2^2\ov x &=& \p_2(\p_2\ov x - \p_1\ov y)-
\p_1\p_0^{-1}\p_1(\p_2\ov x - \p_1\ov y)\\
&=&\p_1\p_2\ov y+\p_1\p_0^{-1}\p_2\p_1\ov x-
    \p_1\p_0^{-1}\p_0\p_2\ov y -
    \p_1\p_0^{-1}\p_2\p_0\ov y\\
&=&\p_1\p_2\ov y - \p_1\p_2\ov y = 0
\end{array}
\]
which completes our analysis of $d_2$.
\renewcommand{\thesection}{Appendix \Alph{section}}
\section{Three lemmas}
\renewcommand{\thesection}{\Alph{section}}
\label{app:l}
\setcounter{equation}{0}
In this appendix we will calculate the cohomology of the following
complex
\begin{equation}
  \label{compl}
  0\to V^{(0)}\onto{d^{(0)}}V^{(1)}\onto{d^{(1)}}V^{(2)}\onto{d^{(2)}}
       V^{(3)}\onto{d^{(3)}}V^{(4)}\onto{d^{(4)}}V^{(5)}\to0,
\end{equation}
where $V^{(0)}\simeq V^{(0)}\simeq\IC[x^{0}\cdots x^{D-1}]$,
$V^{(0)}\simeq V^{(0)}\simeq\IC^{2D}[x^{0}\cdots x^{D-1}]$ and
$V^{(0)}\simeq V^{(0)}\simeq\IC^{D^2+1}[x^{0}\cdots x^{D-1}]$.
Following the notations of section~\ref{s:zclo} we represent the
elements of $V^{(0)}$ and $V^{(5)}$ as $(R^{[0]})$ and $(R^{[5]})$,
elements of $V^{(2)}$ and $V^{(4)}$ as $(Q_{\mu\nu}^{[2]},R^{[2]})$
and $(Q_{\mu\nu}^{[3]},R^{[3]})$, and elements of $V^{(2)}$ and
$V^{(4)}$ by $(P_\mu,\ov P_\mu)$. Differentials $d^{(n)}$ act as
follows
\begin{equation}
  \label{dacts}
  \begin{array}{rcl}
(R^{[0]})                  &\sev{d^{(0)}}&(0)  \\
(P_\mu^{[1]},\; \ov P_\mu^{[1]})
                           &\sev{d^{(1)}}&
                 (\p_\mu R^{[0]},\; \p_\mu R^{[0]})\\
( Q_{\mu\nu}^{[2]},\;R^{[2]})
                           &\sev{d^{(2)}}&
                 ( \p_\mu\ov P_\nu^{[1]} - \p_\nu P_\mu^{[1]},\;
                   \p^\mu\ov P_\mu^{[1]} - \p^\mu P_\mu^{[1]})\\
(Q_{\mu\nu}^{[3]},\;R^{[3]})
                           &\sev{d^{(3)}}&
                 (\dl Q_{\mu\nu}^{[3]},\; \dl R^{[3]})\\
(P_\mu^{[4]},\;\ov P_\mu^{[4]})
                           &\sev{d^{(4)}}&
                 (\p^\nu Q_{\mu\nu}^{[3]}+\p_\mu R^{[3]},
                  \;\p^\nu Q_{\nu\mu}^{[3]}+\p_\mu R^{[3]})\\
(R^{[5]})                  &             &
                 (\p_\mu P^{[4]}-\p_\mu\ov P^{[4]})
  \end{array}
\end{equation}
where
\begin{eqnarray}
  \label{dq}
  \dl Q_{\mu\nu}^{[3]}&=& \Box Q_{\mu\nu}^{[2]}-
     \p^\lm\p_\mu Q_{\lm\nu}^{[2]}-\p^\lm\p_\nu Q_{\mu\lm}^{[2]}+
      2\p_\mu\p_\nu R^{[2]}\\
  \dl R^{[3]}  &=&-\p^\lm\p^\rho Q_{\lm\rho}^{[2]}+
      2\Box R^{[2]}
\end{eqnarray}
It is obvious that $d^{(4)}$ is surjective and the kernel of $d^{(0)}$
contains only constant polynomials. Thus we conclude that $H^0=\IC$
and $H^5=0$. The other cohomology spaces are described by the
following lemmas
\begin{lemma}
  \label{l:h1}
  $H^1$ is finite dimensional and $\dim\,H^1={D(D+1)\over2}$. $H^1$
    cohomology classes can be represented by polynomials of degree no
    bigger than one.
\end{lemma}
{\bf Proof.} According to (\ref{dacts}) $H^1$ is a quotient of the space
$S$ of solutions to the system of first order differential equations
\begin{equation}
  \label{sys}
  \cases{\p_\mu P_\nu=\p_\nu \ov P_\mu&\cr
          \p^\mu P_\mu=\p^\mu \ov P_\mu&}
\end{equation}
by the space $T$ of trivial solutions $P_\mu=\ov P_\mu=\p_\mu R$. Note
that both $S$ and $T$ naturally decompose into direct sum of the
spaces of homogeneous polynomials and so does the quotient
\begin{equation}
  \label{dirs}
  S=\bigoplus S^n,\quad   T=\bigoplus T^n,\quad
  H^1=S/T=\bigoplus S^n/T^n
\end{equation}
We want to prove that $S^n=T^n$ for $n>1$. Let $P_\mu$ and $\ov P_\mu$
be homogeneous polynomials of degree $n>1$ that satisfy \eq{sys}.
First, it is obvious that $P_\mu=0$ for every $\mu$ requires $\ov
P_\mu=0$. Indeed, if $P_\mu=0$ for every $\mu$ then according to the
first equation in \eq{sys} $\p_\nu\ov P_\mu=0$ for every $\mu$ and
$\nu$ and since by assumption $\deg\ov P_\mu>1$ this means $\ov
P_\mu=0$.  Second, using the first equation of (\ref{sys}) twice we
obtain
\begin{equation}
  \label{tws}
  \p_a\p_\nu P_\mu=\p_\a\p_\mu\ov P_\nu=\p_\mu\p_\a\ov P_\nu=
  \p_\a\p_\mu P_\mu.
\end{equation}
Therefore for any $\a$, $\mu$  and $\nu$
\begin{equation}
  \label{amn}
  \p_\mu(\p_\a P_\mu -\p_\mu P_\a)=0.
\end{equation}
And since $\deg(\p_\a P_\mu -\p_\mu P_\a)=n-1>0$ we conclude that
$\p_\a P_\mu -\p_\mu P_\a=0$ and there exist $R$ such that
$P_\mu=\p_\mu R$. Subtracting a trivial solution $(\p_\mu R, \p_\mu
R)$ from $(P_\mu,\ov P_\mu)$ we get another solution $(P'_\mu=0,\ov
P_\mu'=\ov P_\mu-\p_\mu R)$. According to our first observation
$P'_\mu=0$ requires $\ov P'_\mu=0$ and thus $\ov P_\mu=P_\mu=\p_\mu
R$.

It is easy to see that there are exactly ${D(D-1)\over2}$ non-trivial
solutions of degree one and $D$ non-trivial constant solutions which
can be written as
\begin{equation}
  \label{ntone}
  P_\mu=-\ov P_\mu=\xi_{[\mu\nu]}x^\nu,\quad\mbox{and}\quad
  P_\mu=-\ov P_\mu=\mit const
\end{equation}
where $\xi_{[\mu\nu]}$ is an antisymmetric tensor.

\begin{lemma}
  \label{l:h3}
  $H^3=0$
\end{lemma}
{\bf Proof.} This is the most difficult lemma in this work. We have to
show that any solution of the system
\begin{equation}
  \label{req}
  \begin{array}{rcl}
    \p^\nu Q_{\mu\nu}^{[3]}+\p_\mu R^{[3]}&=&0\\
    \p^\nu Q_{\nu\mu}^{[3]}+\p_\mu R^{[3]}&=&0
  \end{array}
\end{equation}
can be represented in the form
\begin{equation}
  \label{trform}
  \begin{array}{rcl}
  \dl Q_{\mu\nu}^{[3]}&=& \Box Q_{\mu\nu}^{[2]}-
     \p^\lm\p_\mu Q_{\lm\nu}^{[2]}-\p^\lm\p_\nu Q_{\mu\lm}^{[2]}+
      2\p_\mu\p_\nu R^{[2]}\\
  \dl R^{[3]}  &=&-\p^\lm\p^\rho Q_{\lm\rho}^{[2]}+
      2\Box R^{[2]}
  \end{array}
\end{equation}
We will start from an arbitrary solution of the system (\ref{req}) and
will be modifying it step by step by adding the trivial solutions of
the form (\ref{trform}) in order to get zero.

We can use the same arguments as in lemma~\ref{l:h1} to consider only
homogeneous polynomials of some degree $m$. Suppose $m\geq1$.  We will
describe an iterative procedure which will allow us to modify
$(Q_{\mu\nu}^{[3]},\; R^{[3]})$ so that $Q_{\mu\nu}^{[3]}$ will depend
only on one variable, say $x^0$ and its only nonzero components be
$Q_{\mu0}^{[3]}$ and $Q_{0\nu}^{[3]}$.

If this is the case, the cocycle condition (\eq{req}) tells us that
\begin{equation}
  \label{h31}
  \p_\mu R =\p_0Q_{\mu,0}=\p_0Q_{0,\mu},
\end{equation}
moreover, $Q_{\mu,0}=C_\mu x_0^m$ and  $Q_{0,\mu}=\ov C_\mu
x_0^m$. Furthermore, using \eq{h31} we conclude that $C_\mu =\ov
C_\mu$. Integrating \eq{h31} we obtain
\begin{equation}
  \label{r3}
  R^{[3]}= \sum_{i=1}^{{D-1}}mC_i x_i x_0^{m-1}+C_0x_0^{26}.
\end{equation}
One can check that such solution $(Q_{\mu\nu}^{[3]},\; R^{[3]})$ can
be written in the form (\ref{trform}) with $Q_{\mu,\nu}^{[2]}=0$ and
\[
R^{[2]}=\half\sum_{i=1}^{{D-1}}\left({C_1\over m+1} x_i x_0^{m+1}+
 {C_0\over (m+1)(m+2)}x_0^{m+2}\right).
\]

Now let us describe the procedure which reduces any solution to the
abovementioned form. Our first objective is to get rid of $x_i$
dependence for $i=1..{D-1}$. Let us pick $i$. The following four step
algorithm will make $(Q_{\mu\nu},R)$ independent of $x_i$. We will
see that when we apply the procedure to $(Q_{\mu\nu},R)$ which does
not depend on some other $x_k$ it will not introduce $x_k$ dependence
in the output.  This observation will allow us to apply the algorithm
${D-1}$ times and make $(Q_{\mu\nu},R)$ depend only on $x_0$.

\begin{itemize}
\item[{\bf Step 1}] Let us introduce some notations. For a polynomial $P$ we
  will denote the minimal degree of $x_i$ among all the monomials in
  $P$ by $n_i(P)$. For a zero polynomial we formally set
  $n_i(0)=+\infty$. Given a matrix of polynomials $Q_{\mu,\nu}$, let
  \begin{equation}
    \label{ndef}
    N_i(Q_{\mu\nu})=
     \min_{\mu\neq i\atop\nu\neq i} n_i(Q_{\mu\nu})
  \end{equation}
  Since $Q_{\mu\nu}^{[3]}$ are homogeneous polynomials of degree $m$
  we can write them as
  \begin{equation}
    \label{qmnc}
    Q_{\mu\nu}^{[3]}=\sum_{m_0+\cdots+m_{{D-1}}=m}{C_{m_0\cdots
        m_{{D-1}},\,\mu\nu}}x_0^{m_0}\cdots x_{{D-1}}^{m_{{D-1}}}
  \end{equation}
  Let us show that it is possible to add a trivial solution to
  $(Q_{\mu\nu},R)$ and increase $ N_i(Q_{\mu\nu})$ by one. Indeed, let
  \[
  Q_{\mu\nu}^{[2]}=\cases{\ds
    \sum_{m_0+\cdots+m_{{D-1}}=m}\kern-25pt-{C_{m_0\cdots m_{{D-1}},\,\mu\nu}\;
      x^{m_0}_{0}\cdots x_i^{m_i+2}\cdots x^{m_{{D-1}}}_{{{D-1}}}\over
      (m_i+1)(m_i+2)}&for $\mu,\nu\neq i$,\cr\cr
    \hfill0\hfill&otherwise,\cr}
  \]
  and $R^{[2]}=0$.

  It is easy to see that
  \[
  N_i(Q_{\mu\nu}^{[3]}+\dl Q_{\mu\nu}^{[3]})\geq N_i(Q_{\mu\nu}^{[3]})+1
  \]
  where $\dl Q_{\mu\nu}^{[3]}$ comes from the trivial solution
  generated by $(Q_{\mu\nu}^{[2]},R^{[2]})$ according to
  (\ref{trform}). Repeating this procedure, we will increase
  $N_i(Q_{\mu\nu}^{[3]})$ at least by one every time. Since
  $Q_{\mu\nu}^{[3]}$ are homogeneous polynomials of degree $m$, $N_i$
  is either less then $m+1$ or equal $+\infty$.  Therefore after a
  finite number of steps we will make $N_i=+\infty$ which means that
  all $Q_{\mu\nu}^{[3]}$ are zero for $\mu\neq i$ and $\nu\neq i$.
\item[{\bf Step 2}] Since $(Q_{\mu\nu}^{[3]},R^{[3]})$ is a solution
  to the system (\ref{req}) we can write
  \[
  \begin{array}{rcl}
    \p_iQ_{\mu i}^{[3]}&=&\p_\mu R^{[3]}=\p_iQ_{i\mu}^{[3]},
    \quad\mu\neq i\\
    \p^\nu Q_{\nu i}^{[3]}&=&\p_i R^{[3]}=\p^\nu Q_{i\nu}^{[3]}.
  \end{array}
  \]
  Suppose
  \[R^{[3]}=\sum_{m_0+\cdots +m_{{D-1}}=m}D_{m_0\cdots
    m_{{D-1}}}x_0^{m_0}\cdots x_{{D-1}}^{m_{{D-1}}},
  \]
  then we choose $R^{[2]}=0$,
  \[Q_{ii}^{[2]}=\sum_{m_0+\cdots +m_{{D-1}}=m}
  {D_{m_0\cdots m_{{D-1}}}\over (m_i+1)(m_i+2)}
  x_0^{m_0}\cdots x_i^{m_i+2}\cdots x_{{D-1}}^{m_{{D-1}}},
  \]
  and $Q_{\mu\nu}^{[2]}=0$ for all the other $\mu$ and $\nu$. It is
  easy to see that $\tQ_{\mu\nu}^{[3]}=Q_{\mu\nu}^{[3]}+\dl
  Q_{\mu\nu}^{[3]}$ have the following properties:
  \begin{itemize}
  \item all  $\tQ_{\mu\nu}^{[3]}$ except $\tQ_{ii}$ do not depend on
    $x_i$
  \item  $\tQ_{\mu\nu}^{[3]}=0$ for all $\mu\neq i$ and $\nu\neq i$.
  \end{itemize}
\item[{\bf Step 3}] Suppose $(Q_{\mu\nu}^{[3]}, R^{[3]})$ is of the
  form we obtained at the end of step~2. Since $\p_i Q_{\mu
    i}^{[3]}=0$ for $\mu\neq i$, then $\p_\mu R^{[3]}=0$ for $\mu\neq
  i$ and therefore, $R^{[3]}$ depends only on $x_i$. Thus there exists
  $R^{[2]}$ such that $R^{[3]}=-2\p_i^2R^{[2]}$ and $R^{[2]}$ depends
  only on $x_i$. adding a trivial solution generated by $(0, R^{[2]})$
  we can make $R^{[3]}=0$.
\item[{\bf Step 4}] Using $R^{[3]}=0$ we can rewrite the system
  (\ref{req}) as follows.
  \[
  \p_i Q_{ii}^{[3]}=-\sum_{\nu\neq i}\p^\nu Q_{\nu i}^{[3]}.
  \]
  Therefore, $ Q_{ii}^{[3]} = Q_{ii,\,0}^{[3]}+x_i Q_{ii,\,1}^{[3]}$,
  where $Q_{ii,\,0}^{[3]}$ and $Q_{ii,\,1}^{[3]}$ do not depend on
  $x_i$. For every $Q_{ii,\,1}^{[3]}$ we can find a polynomial $P$
  which depends on the same set of variables and $\Box P=-
  Q_{ii,\,1}^{[3]}$. Choose $Q_{ii}^{[2]}=x_i\,P$, $R^{[2]}=0$ and
  $Q_{\mu\nu}^{[2]}=0$ for $(\mu,\nu)\neq(i,i)$. Adding the
  corresponding trivial solution we achieve that $Q_{\mu\nu}^{[3]}$
  and $R^{[3]}$ do not depend on $x_i$.
\end{itemize}
Repeating this program ${D-1}$ times for each value of $i=1,\dots,{D-1}$, we
make $(Q_{\mu\nu}^{[3]}, R^{[3]})$ depend only on $x_0$. Now we can
repeat the first step once again with $i=0$ and make $Q_{kj}^{[3]}=0$
for $k,j=1,\dots {D-1}$. We have already proven that such solution is
trivial.

Recall that in the very beginning of our analysis we have made an
assumption that the polynomials have non-zero degree ($m\geq1$).
Therefore we have to consider this last case separately. If
polynomials $Q_{\mu\nu}^{[3]}$ and $R^{[3]}$ are constant, they
trivially satisfy the system~(\ref{req}). To show that any such
constant solution can be represented in the form (\ref{trform}), it is
sufficient to take
$Q^{[2]}_{\mu\nu}=q_{\mu\nu,0}(x_0)^2/2+q_{\mu\nu,1}(x_1)^2/2$, which
generates $(Q_{\mu\nu}^{[3]},R^{[3]})$ if $q_{\mu\nu,0}$ and
$q_{\mu\nu,1}$ are chosen so that
\begin{eqnarray*}
  q_{00,1}-q_{00,0}&=&Q_{00}^{[3]},\\
  q_{11,0}-q_{11,0}&=&Q_{11}^{[3]},\\
  q_{00,1}+q_{00,0}+q_{11,1}+q_{11,0}&=&R^{[3]},\\
  q_{0\nu,1}&=&Q_{0\nu}^{[3]}\mbox{ for $\nu\neq0$},\\
  q_{\mu0,1}&=&Q_{\mu0}^{[3]}\mbox{ for $\mu\neq0$},\\
  q_{1\nu,0}&=&Q_{1\nu}^{[3]}\mbox{ for $\nu\neq1$},\\
  q_{\mu1,0}&=&Q_{\mu1}^{[3]}\mbox{ for $\mu\neq1$},\\
  q_{\mu\nu,0}+q_{\mu\nu,1}&=&Q_{\mu\nu}^{[3]}\mbox{ for $\mu,\nu>1$}.
\end{eqnarray*}
This completes the proof of Lemma~\ref{l:h3}.

\begin{lemma}
  \label{l:h4}
  $H^4=0$.
\end{lemma}
{\bf Proof.} It is almost obvious that the image of $d^{(3)}$ covers
the whole kernel of $d^{(4)}$ in $V^{(4)}$ because the space $V^{(3)}$
is much bigger than $V^{(4)}$ at every degree. Indeed we will show
that it is sufficient to consider a subspace of $V^{(3)}$ spanned by
the zeroth row and the zeroth column of the matrix $Q_{\mu\nu}$.
Loosely speaking the row will cover $P_\mu$ and the column will cover
$\ov P_\mu$.

Suppose $(P_\mu,\ov P_\mu)\in \ker{d^{(4)}}$ or equivalently
\begin{equation}
  \label{d4cond}
  d^{(4)}(P_\mu,\ov P_\mu)=(\p^\mu P_\mu-\p^\mu\ov P_\mu)=0
\end{equation}
we want to show that subtracting vectors of the form
$d^{(3)}(Q_{\mu\nu}, 0)$ from $(P_\mu,\ov P_\mu)$ we can get reduce it
to zero. First of all we can easily get rid of the spacial components
$P_i$ and $\ov P_i$ for $i=1\cdots D-1$ using $Q_{\mu\nu}$ which has
the only non-zero components given by
\begin{equation}
  \label{ais}
   Q_{0i}=\int P_i\d x^0 \quad\mbox{and}\quad
   Q_{i0}=\int \ov P_i\d x^0  + C_i(x^1\cdots x^{D-1})
\end{equation}
This will reduce $P_\mu$ and $\ov P_\mu$ to the form
$P_\mu=a\dl_{0,\mu}$ and $\ov P_\mu=\ov a\dl_{0,\mu}$. Furthermore,
according to \eq{d4cond} polynomials $a$ and $\ov a$ have the same
derivative with respect to $x^0$. Thus varying say $C_1(x^1\cdots
x^{D-1})$ in \eq{ais} we can achieve that $a=\ov a$. Finally if we
have a vector given by $P_\mu=\ov P_\mu=a\dl_{0,\mu}$ we can use
$Q_{00}$ to reduce it to zero.
%\bibliography{personal}
%\bibliographystyle{unsrt}
%\end{document}

\end{document}